\documentclass[twocolumn]{revtex4-2}
\usepackage{graphicx}
\usepackage{amsmath}
\usepackage{amssymb}
\usepackage{bm}
\usepackage{color}
\usepackage{hyperref}
\usepackage{tabularx}
\usepackage{multirow}
\usepackage{braket}
\usepackage{epstopdf}

\makeatletter
\def\blfootnote{\gdef\@thefnmark{}\@footnotetext}
\makeatother

\begin{document}

\title{
Magic in twisted transition metal dichalcogenide bilayers  
}

\author{Trithep Devakul$^{1*\dagger}$
}
\author{Valentin Cr\'epel$^{1}$}
\author{Yang Zhang$^{1\dagger}$}
\author{Liang Fu$^{1*}$}
\affiliation{$^1$ Department of Physics, Massachusetts Institute of Technology, Cambridge, Massachusetts 02139, USA}

\blfootnote{$^\dagger$ These authors contributed equally: Trithep Devakul, Yang Zhang.}
\blfootnote{$^*$ Correspondence should be addressed to Trithep Devakul (email:tdevakul@mit.edu) or Liang Fu (email:liangfu@mit.edu).}

\begin{abstract}
The long wavelength moir\'e superlattices in twisted 2D structures have emerged as a highly tunable platform for strongly correlated electron physics.
We study the moir\'e bands in twisted transition metal dichalcogenide homobilayers, focusing on WSe$_2$, at small twist angles using a combination of first principles density functional theory, continuum modeling, and Hartree-Fock approximation.
We reveal the rich physics at small twist angles $\theta<4^\circ$, and identify a particular magic angle at which the top valence moir\'e band achieves almost perfect flatness.
In the vicinity of this magic angle, we predict the realization of a generalized Kane-Mele model with a topological flat band, interaction-driven Haldane insulator, and Mott insulators at the filling of one hole per moir\'e unit cell.
The combination of flat dispersion and uniformity of Berry curvature near the magic angle holds promise for 
realizing fractional quantum anomalous Hall effect at fractional filling. We also identify twist angles favorable for 
quantum spin Hall insulators and interaction-induced quantum anomalous Hall insulators 
at other integer fillings.  

\end{abstract}
                          	
\maketitle
\section{Introduction}
In condensed matter physics, simple and elegant models have often brought new ideas and started new paradigms. 
Celebrated examples include the Hubbard model for strongly correlated electron system~\cite{Hubbard}, 
the Tomonaga-Luttinger model for one-dimensional electron liquid~\cite{Tomonaga,Luttinger}, 
and the Kitaev model for non-Abelian quantum spin liquid~\cite{Kitaev}, to name a few. 
As toy models are designed to illustrate key concepts in the simplest form, 
they are rarely realized directly in real materials, whose atomic-scale electronic structures are inevitably more complex.  
The recent advent of { long-wavelength} moir\'e superlattices based on 2D van der Waals structures provides a new and promising venue for physical realization and quantum simulation of model Hamiltonians.   
In magic-angle twisted bilayer graphene~\cite{Bistritzer2011} (TBG), 
experiments have discovered a variety of correlated electron states~\cite{Cao2018, Cao2018a,Yankowitz2019, Lu2019,Cao2020, Sharpe2019,Serlin2020} 
facilitated by flat moir\'e bands. 

More recently, moir\'e superlattices of semiconducting transition metal dichalcogenides (TMD) 
have attracted interest as a potentially simpler and more robust platform for simulating the Hubbard model 
on an emergent lattice ~\cite{Wu2018,Wu2019,Tang2020,Regan2020,Shabani2021,Jin2021,Zhang2020,Zhang2021,Slagle2020,Xu2020,Bi2021,Pan2020,MoralesDuran2020,Zang2021,padhi2021generalized,zhai2020theory,Cazalilla2014,Zhang2020flat,magorrian2021multifaceted,tang2021geometric}. Each lattice site represents a { low-energy} electronic orbital in the moir\'e unit cell  that spreads over { many} atoms.  These semiconductor moir\'e systems can thus be viewed as artificial 2D solids---a periodic array of
``magnified atoms'' \cite{Zhang2020}.  The atomic potential depth and interatomic bonding are highly tunable by the choice of TMD materials, the twist angle \cite{Wu2018} and the displacement field \cite{Bi2021, Zhang2021}. 
Thus, TMD based moir\'e materials provide a favorable platform for simulating idealized models in two dimensions.

In this work, we predict the realization of generalized Kane-Mele models with topological flat  band, 
interaction-driven Haldane insulator and Mott insulators in twisted TMD homobilayers at small twist angles.  
Contrary to current thoughts, we show by band structure calculation and analytical derivation that a magic twist angle exists in twisted TMD homobilayers, where the topmost valence miniband from the $\pm K$-valleys is almost perfectly flat and well separated from other bands. This band carries a spin/valley Chern number and is well described by a generalized Kane-Mele model~\cite{Kane2005}.

At half filling of this topological flat band, we show that repulsive interactions 
drives spontaneous spin/valley polarization leading to Haldane's quantum anomalous Hall insulator~\cite{Haldane1988}.
We further find an out-of-plane displacement field drives a transition from the Haldane insulator into a Mott insulator. Depending on the twist angle, this Mott state is either a spin/valley polarized ferromagnet or features intervalley coherence that spontaneously breaks the spin/valley $U(1)$ symmetry. 
Thus our work reveals a rich phase diagram of topological, correlated and broken-symmetry insulators 
enabled by the flat band in TMD homobilayers at small twist angles below the $4^\circ$--$5^\circ$ range in current experimental studies ~\cite{Wang2020, Bi2021}.  

Due to spin-valley locking~\cite{Di2012}, 
monolayer TMDs such as  WSe$_2$ and MoTe$_2$ feature top valence bands with spin-$\uparrow$ at $+K$ valley and spin-$\downarrow$ at $-K$. We study TMD homobilayers with a small twist angle $\theta$ starting from AA stacking, where every metal (M) or chalcogen (X) atom  on  the top layer is aligned with the same type of atom on the bottom layer. 
In such twisted structure, the $K$ points of the two layers are slightly displaced and form the two corners of the moir\'e Brillouin zone, denoted as $\kappa_\pm$. 
A set of spin-$\uparrow$ ($\downarrow$) moir\'e bands is formed from hybridized $+K$ ($-K$) valley bands of the two layers. 
The complete filling of a single moir\'e band including spin degeneracy thus requires 2 holes per moir\'e unit cell. 

\section{Results}  

\begin{figure}[t]
\includegraphics[width=\columnwidth]{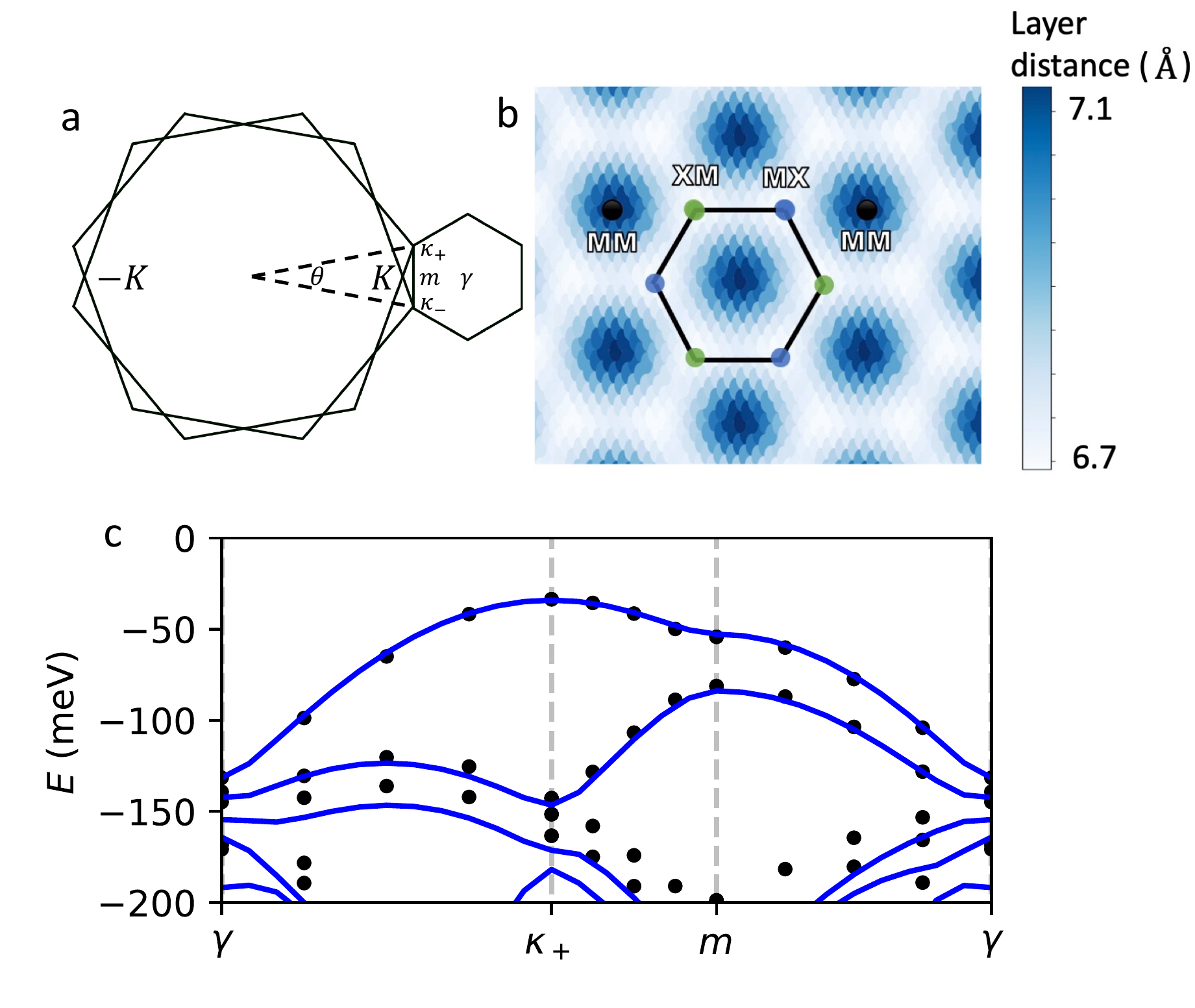}
\caption{
{ Comparison with large scale DFT calculations.}
$a)$ The $\kappa_{\pm}$ points of the moir\'e Brillouin zone are formed from the $K$ points of the monolayer Brillouin zones, which are rotated by $\pm \theta/2$.
$b)$ The interlayer distance of the twisted WSe$_2$ structure obtained from DFT is shown, demonstrating a large variation between the MM and XM/MX regions.
$c)$ The continuum band structure (blue lines) is plotted in comparison with large scale DFT calculations (black dots) at twist angle $\theta=5.08^\circ$, showing excellent agreement.
}\label{fig:bzdft}
\end{figure}

In order to obtain accurate moir\'e band structures, we perform large-scale density functional theory calculations with the SCAN+rVV10 van der Waals density functional ~\cite{peng2016versatile}, which captures the intermediate-range vdW interaction through its semilocal exchange term. 
Focusing on twisted bilayer WSe$_2$, we find that lattice relaxation has a dramatic effect on moir\'e bands.  
Our DFT calculations at $\theta=5.08^\circ$ with 762 atoms per unit cell show a significant variation of the layer distance $d$ in different regions on the moir\'e superlattice, as shown in Fig~\ref{fig:bzdft}b.  $d=6.7$\AA{} is smallest in MX and XM stacking regions, where the metal atom on top layer is aligned with chalcogen atom on the bottom layer and vice versa, while $d=7.1$\AA{} is largest in MM region where metal atoms of both layers are aligned. With the fully relaxed structure, the low-energy moir\'e valence bands of twisted bilayer WSe$_2$ are found to come from the $\pm K$ valley (shown in Fig.1c), as opposed to 
the $\Gamma$ valley in previous computational studies~\cite{naik2018ultraflatbands} 
and consistent with recent works \cite{Wang2020,vitale2021flat,kundu2021flat}. 

At small twist angles, the large size of moir\'e unit cell makes it difficult to perform DFT calculations directly on twisted TMD homobilayers. An alternative and complementary approach, introduced by Wu et al. \cite{Wu2019}, is the continuum model based on an effective mass description, which models the formation of moir\'e bands using spatially-modulated interlayer tunneling $\Delta_T(\textbf{r})$  and layer-dependent potential $\Delta_{1,2}(\textbf{r})$. 
The continuum model Hamiltonian  for $\pm K$ valley bands is given by
\begin{equation}
\mathcal{H}_{\uparrow} = 
\begin{pmatrix}
-\frac{\hbar^2 (\textbf{k}-\bm{\kappa}_+)^2}{2 m^*} + \Delta_1(\textbf{r}) & \Delta_T(\textbf{r})\\
 \Delta_T^\dagger(\textbf{r})
 &-\frac{\hbar^2 (\textbf{k}-\bm{\kappa}_-)^2}{2 m^*} + \Delta_2(\textbf{r})
\end{pmatrix}
\label{eq:Hcont}
\end{equation}
and $\mathcal{H}_{\downarrow}$ as its time-reversal conjugate. 

The continuum model approach is valid at small twist angle where the moir\'e wavelength is large enough.  
In this case, the atom configuration within any local region of a twisted bilayer is identical to that of an untwisted bilayer 
with one layer laterally shifted relative to the other by a corresponding displacement vector ${\textbf{d}}_0$. For example, ${\textbf{d}}_0=0, -\left(\textbf{a}_{1}+{\textbf{a}}_{2}\right) /3, \left(\textbf{a}_{1}+{\textbf{a}}_{2}\right) /3$, with ${\textbf{a}}_{1,2}$ the primitive lattice vector of a monolayer, correspond to the MM, MX and XM regions respectively.  
Therefore the moir\'e potentials for twisted TMD bilayers $\Delta_T(\textbf{r})$ and $\Delta_{1,2}(\textbf{r})$ as a function of 
coordinate on the moir\'e superlattice can be determined from the valence band edges
of the untwisted bilayer as a function of the corresponding shift vector \cite{Zhang2021}.
In the lowest harmonic approximation, $\Delta_T(\textbf{r})$ and $\Delta_{1,2}(\textbf{r})$ are sinusoids that interpolate between MM, MX and XM regions~\cite{Wu2019}: 
\begin{eqnarray}
\Delta_{1,2}(\textbf{r}) &=& 2 V \sum_{j=1,3,5} \cos(\textbf{g}_j\cdot \textbf{r} \pm \psi)\\
\Delta_{T}(\textbf{r}) &=& w (1 + e^{-i \textbf{g}_2\cdot\textbf{r}} + e^{-i \textbf{g}_3\cdot\textbf{r}})
\end{eqnarray} 
where $\textbf{g}_j$ are $(j-1)\pi/3$ counter-clockwise rotations of the moir\'e reciprocal lattice vector $\textbf{g}_1=(4\pi\theta/\sqrt{3}a_0,0)$, and $a_0$ is the monolayer lattice constant.
Up to an overall energy scale, the continuum model depends only on the dimensionless parameters $\alpha \equiv V \theta^2/(m^* a_0^2)$, $w/V$ and $\psi$.

From our DFT calculation for untwisted bilayers with relaxed layer distance and using the vacuum level 
as an absolute reference energy for the band edge, we obtain the continuum model parameters $V=9.0$meV, $\psi=128^\circ$ and $w=18$meV. 
Importantly, the interlayer tunneling strength $w$ is twice larger than previously reported \cite{Wu2019}.
To demonstrate the accuracy of the continuum model method, 
we compare 
in Fig~\ref{fig:bzdft}c the band structures 
computed by large-scale DFT directly at $\theta=5.08^\circ$ and by the continuum model 
with the above parameters,  
finding excellent agreement.  
{ Details on the DFT calculation can be found in Supplementary Note 1~\cite{supp}.}

We remark that different approaches~\cite{magorrian2021multifaceted,tang2021geometric,kundu2021flat} can lead to different conclusions on topology.  
Thus, we utilize a method to determine band topology directly from our large-scale DFT band structure based on symmetry eigenvalues.
{ As detailed in Supplementary Note 2~\cite{supp}}, we are able to isolate bands from the $\pm K$ valley and compute their $C_{3z}$ eigenvalue at the high symmetry momenta $\gamma$, $\kappa_{\pm}$,
which determine their Chern number (mod 3)~\cite{Fang2012}.
The $C_{3z}$ eigenvalues for the first two bands, summarized in Table~\ref{table1}, are consistent with the first two bands having non-trivial valley Chern number $\mathcal{C}_{K,1}=\mathcal{C}_{K,2}=1$.

\begin{table}[h]
\centering
\begin{tabular}{|l|l|l|l|}
\hline
Band, Valley &	$\kappa_+$	& $\kappa_-$	& $\gamma$\\
\hline
1, $K$   & $e^{i\pi/3}$	& $e^{i\pi/3}$ & $e^{i\pi}$ \\
1, $K^\prime$ & $e^{-i\pi/3}$ & $e^{-i\pi/3}$ & $e^{i\pi}$ \\
2, $K$   & $e^{-i\pi/3}$ & $e^{-i\pi/3}$ & $e^{i\pi/3}$ \\
2, $K^\prime$ & $e^{i\pi/3}$ & $e^{i\pi/3}$ & $e^{-i\pi/3}$ \\
\hline
\end{tabular}
\caption{$C_{3z}$ eigenvalues of the first two bands from each valley, computed from large-scale DFT wavefunctions at high symmetry momentum points.}\label{table1}
\end{table}

\begin{figure}[t]
\includegraphics[width=\columnwidth]{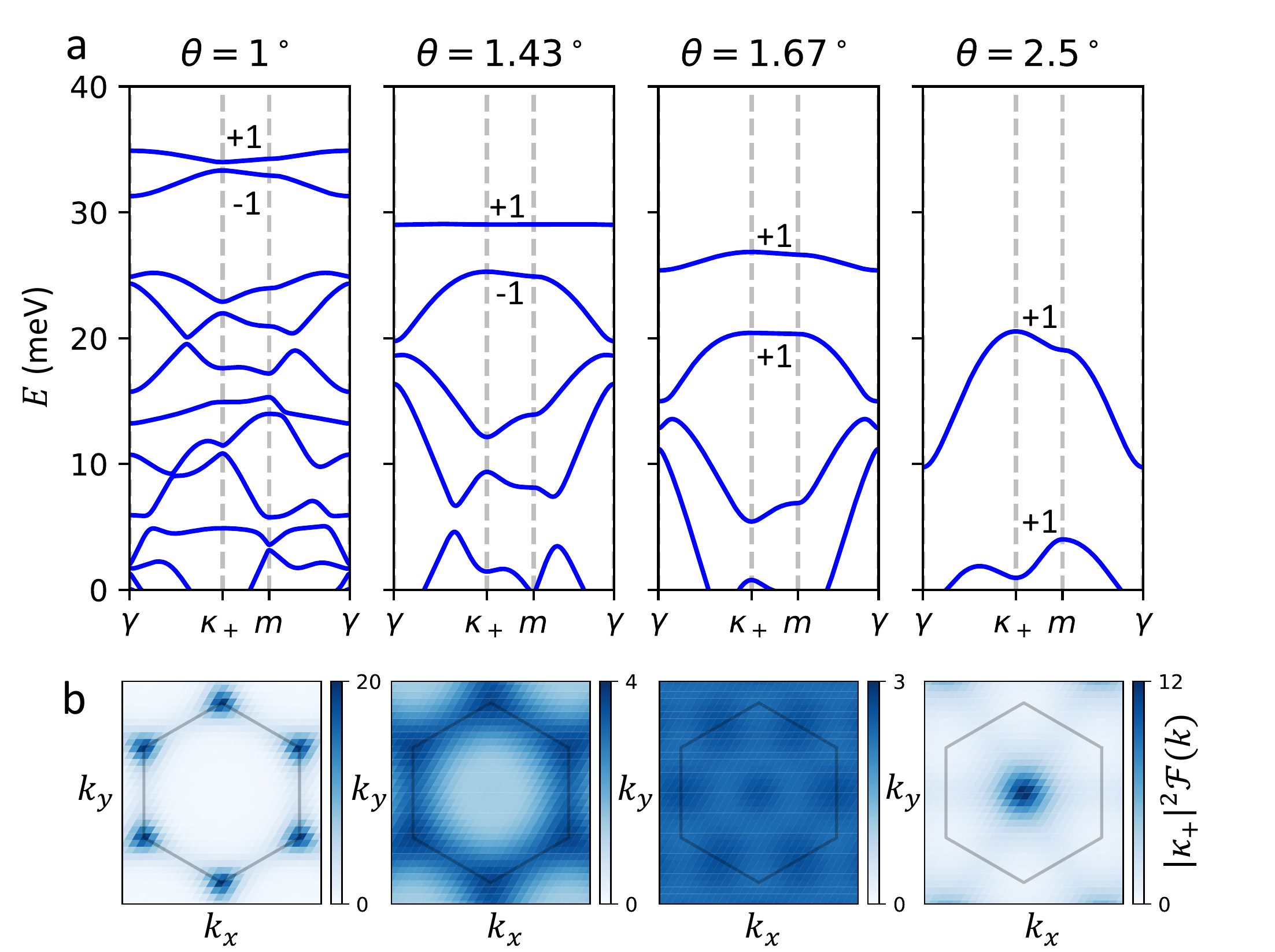}
\caption{
{ Continuum model band structure and Berry curvature at various twist angles.}
$a)$ The band structure $E_i(\textbf{k})$ along with the Chern numbers of the first two bands and $b)$ the (scaled) Berry curvature $|\kappa_+|^2\mathcal{F}(\textbf{k})$ of the first band is shown for the continuum model at $\theta = 1^\circ, 1.43^\circ,1.67^\circ,$ and $2.5^\circ$.
At $\theta=2.5^\circ$, the first band maxima is located at the $\kappa_{\pm}$ points and the Berry curvature is peaked at $\gamma$.
At $\theta=1^\circ$, the band maximum is instead at $\gamma$ and $\mathcal{F}$ is peaked at $\kappa_{\pm}$.
During the crossover region between these angles, $E_1(\textbf{k})$ and $\mathcal{F}(\textbf{k})$ both become very flat.
We find that the band dispersion $E_1(\textbf{k})$ is flattest at $\theta\approx1.43^\circ$ and the Berry curvature $\mathcal{F}(\textbf{k})$ is most uniform at $\theta\approx1.67^\circ$, both shown.
}\label{fig:BandBerry}
\end{figure}

\begin{figure}[t]
\includegraphics[width=\columnwidth]{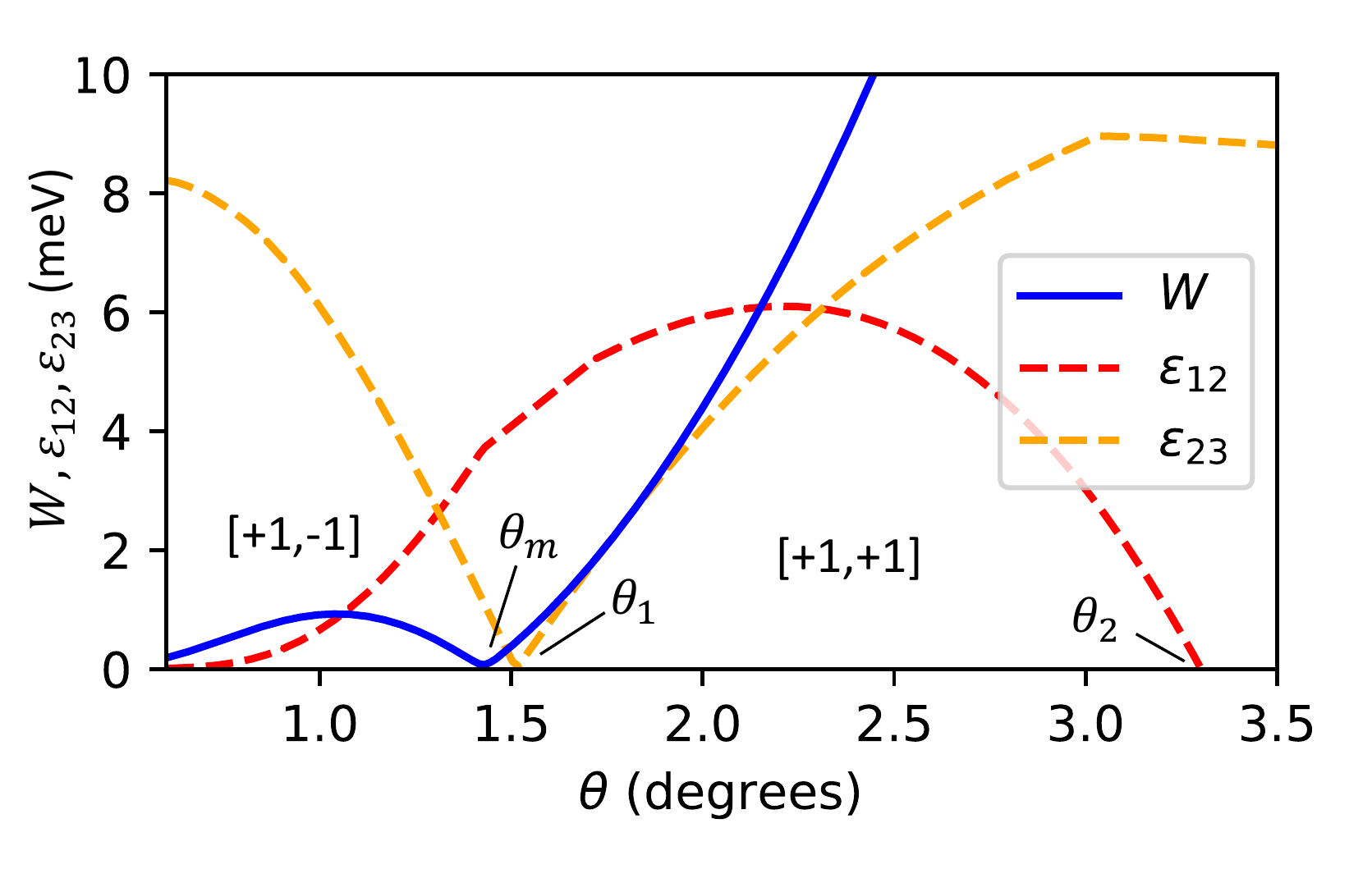}
\caption{The bandwidth of the first band $W$, and indirect band gap between the first two pairs of bands $\varepsilon_{12}$ and $\varepsilon_{23}$.  
The bandwidth is minimized at $\theta_m$, while being well separated from the remaining bands.
The Chern numbers of the first two bands, $[\mathcal{C}_{K,1},\mathcal{C}_{K,2}]$, is shown before and after the $\varepsilon_{23}$ gap closing at $\theta=\theta_1$.
For $\theta\geq\theta_2$, $\varepsilon_{12}$ vanishes.
}\label{fig:gaps}
\end{figure}

Using the new continuum model parameters established above,
along with the lattice constant $a_0 = 3.317\text{\AA{}}$~\cite{mounet2018two} and the effective mass $m^* = 0.43 m_e$ \cite{fallahazad2016shubnikov,rasmussen2015computational}, 
we calculate the band structure of twisted bilayer WSe$_2$, $E_i(\textbf{k})$, at various twist angles, as shown in 
Figure~\ref{fig:BandBerry}a.
The bandwidth of the first band,
$W = \max_{\textbf{k}} E_1(\textbf{k}) - \min_{\textbf{k}} E_1(\textbf{k})$,
as well as the (direct or indirect) band gaps $\varepsilon_{ij}$ between pairs of bands $(i,j)=(1,2)$ and $(2,3)$,  $\varepsilon_{ij} = \min_{\textbf{k}} E_i(\textbf{k}) - \max_{\textbf{k}} E_j(\textbf{k})$,
is shown in Figure~\ref{fig:gaps}.
Focusing on topological features of the first two valence bands, we can divide the moir\'e band structure into three main regimes divided by
$\theta_1\approx 1.5^\circ$ and $\theta_2 \approx 3.3^\circ$.

First, for $\theta < \theta_1$,
the top two bands are well separated from the rest of the spectrum, and carry opposite Chern number $[\mathcal{C}_{K,1},\mathcal{C}_{K,2}] = [+1,-1]$.
The bandwidth of the first band $W < 1\text{meV}$ remains very small throughout.
In this regime of very small twist angles, the character of the top two valence bands can be understood from an effective tight binding model on a moir\'e honeycomb lattice that takes the form of a Kane-Mele model, as suggested in the insightful work of Wu et al.~\cite{Wu2019}
As we will later show, the original Kane-Mele description with up to second nearest neighbor hopping terms 
only describes the band structure well for very small angles $\theta \lesssim 1^\circ$.
As $\theta$ increases towards $\theta_1$, longer range hopping terms become more important.

At $\theta=\theta_1$, the band gap $\varepsilon_{23}$ closes and the Chern number of the top two bands changes to $[+1,+1]$.
In this second regime, $\theta_1 < \theta < \theta_2$, both top bands have same Chern number $[+1,+1]$ and are still all separated by a sizable gap $\varepsilon_{12},\varepsilon_{23}>0$.  
The bandwidth of the first band increases rapidly with $\theta$, reaching around $W \approx 20\text{meV}$ at $\theta_2$ (not shown). 
Finally, in the third regime, $\theta_2<\theta\lesssim 5.4^\circ$, the indirect gap $\varepsilon_{12}$ vanishes, but the direct gap remains open.
The Chern number of the top two bands remains well defined at $[+1,+1]$, but the bands now overlap in energy and are highly dispersive.
In both the second and third regimes, $\varepsilon_{23}>0$, thus the top two bands together form a gapped $\mathcal{C}=2$ manifold. 
Beyond $\theta \gtrsim 5.4^\circ$, the gap $\varepsilon_{23}$ also vanishes and the top two bands are no longer isolated { (Supplementary Note 3)\cite{supp}}.
 Topology of the continuum model at $\theta\approx 5^\circ$ is consistent with that determined directly from large-scale DFT in Table~\ref{table1}, further strengthening our confidence in the continuum model description even up to larger angles.

For $\theta < \theta_2$, especially near $\theta_2$ where the first band is more dispersive,
the spin Chern number $\mathcal{C}=1$ and $\varepsilon_{12}>0$ is favorable for a quantum spin Hall insulator at a filling of $n=2$ holes per moir\'e unit cell.
Also, for the wide range of angles $\theta_1<\theta\lesssim 5.4^\circ$, $\varepsilon_{23}>0$ and the top two bands both carry spin Chern number $\mathcal{C}=1$, giving rise to
 a double quantum spin Hall state with two sets of counter-propagating spin-polarized edge modes at $n=4$.

We now address the bandwidth $W$, which shows a sharp minimum at $\theta=\theta_m \approx 1.43^\circ$ reminiscent of the magic angle in TBG.
To understand this, notice that the top band, shown in Figure~\ref{fig:BandBerry}a, has two qualitatively different behaviors in the small and large $\theta$ limit.
For $\theta\gtrsim 2.5^\circ$, $E_1(\textbf{k})$ reaches its maximum at $\bm{\kappa}_{\pm}$ and minimum at $\gamma$, which can be understood from the weak moir\'e effects at small $\alpha$.
For small $\theta\lesssim 1^\circ$, the opposite holds and $E_1(\textbf{k})$ is maximum is at $\gamma$ and minimum at $\kappa_{\pm}$, which can be understood from the effective Kane-Mele model, which we will derive explicitly.
At the crossover between these two limits, the band maxima and minima must switch locations in the moir\'e Brillouin zone, potentially leading to a flat band.
As can be clearly seen, $W$ achieves a minimum at a particular magic angle $\theta_m$ during this crossover.
At $\theta_m$ the gap to the next state $\varepsilon_{12}\approx3.7\text{meV}$ is much larger than the bandwidth $W \approx 0.1\text{meV}$.
The band structure at $\theta_m$ is shown in Fig~\ref{fig:BandBerry}a, which shows that the first band is almost completely flat and separated from the next band.
For even smaller $\theta$, both $\varepsilon_{12}$ and $W$ vanish, but the ratio $W/\varepsilon_{12}$ diverges. 
Thus, we may view $\theta_m$ as the angle at which the top band is flattest while still being well isolated from the rest of the spectrum.

Analytic progress can be made in estimating the magic angle by considering the dispersion near $\bm{\gamma}$.
Assuming that the bandwidth will be minimized near the angle at which $E_1(\bm{\gamma})$ changes from minima to maxima,
expanding $E_1(\bm{\gamma}+\textbf{k}) \approx E_1(\bm{\gamma}) + \frac{\textbf{k}^2}{2 m_\gamma} + \mathcal{O}(\textbf{k}^3)$,
the effective mass $m_\gamma$ should diverge near the crossover.
Let $\tilde{\theta}_m$ to be the angle at which $m_\gamma^{-1}=0$.
Then, considering only the $6$ most relevant states at $\bm{\gamma}$, we have { (Supplementary Note 4~\cite{supp})}
\begin{equation}
\tilde{\theta}_m^{-2} = \frac{8\pi^2}{9 m^* a_0^2} \left(\frac{1}{\mathcal{E}_{n_0} - \mathcal{E}_{n_0+1}} + \frac{1}{\mathcal{E}_{n_0} - \mathcal{E}_{n_0-1}}\right)
\label{eq:tildetheta}
\end{equation}
where $\mathcal{E}_{n} = 2 w \cos (\pi n / 3) + 2 V \cos(2 \pi n / 3 - \psi)$, and $n_0$ is the integer (mod 6) which maximizes $\mathcal{E}_n$ ($n_0=1$ for WSe$_2$ parameters).
We find that Eq~\eqref{eq:tildetheta} provides a decent estimate for $\theta_m$ in the cases considered.
In WSe$_2$, we have $\tilde{\theta}_m = 1.47^\circ$, compared to $\theta_m=1.43^\circ$ at which the bandwidth is minimized.

Next, we turn to the Berry curvature $\mathcal{F}(\textbf{k})$ of the top band, shown in Fig~\ref{fig:BandBerry}b.
In all cases, the first band has Chern number $\mathcal{C}_{K,1}=\frac{1}{2\pi}\int_{\mathrm{BZ}}\mathcal{F} d \textbf{k} = 1$; however, the distribution changes very drastically as $\theta$ is varied.
At $\theta\gtrsim2^\circ$, $\mathcal{F}$ is peaked around the band minimum at $\bm{\gamma}$.
At $\theta\lesssim 1^\circ$, $\mathcal{F}$ is instead sharply peaked at the $\bm{\kappa}_{\pm}$ points.
Near the crossover region, the distribution of $\mathcal{F}$ shifts from $\bm{\gamma}$ to $\bm{\kappa}_{\pm}$, and can become very evenly distributed.
We find that $\mathcal{F}$ is most evenly distributed near $\theta=1.67^\circ$, shown in Fig~\ref{fig:BandBerry}b, where $\mathcal{F}$ becomes almost uniform in the Brillouin zone.
The uniform distribution of $\mathcal{F}$ is reminiscent to that of Landau levels.
Time reversal symmetry forces the corresponding spin-$\downarrow$ bands from the $-K$ valley to have opposite Chern number. 

We emphasize that the physics of the magic angle arises due to the crossover between two qualitatively different behaviors of the first band at low and high angles.
Additional factors unaccounted for by the continuum model may result in, for example, angle-dependent model parameters.
However, as long as the qualitative behaviors at small and large angles are unchanged, there will still be crossover regime at which the band becomes flat. 
Even when the bands are not perfectly flat, a diverging mass can still give rise to a diverging higher-order van Hove singularity~\cite{Bi2021}.

Recall that for $\theta < \theta_1$, the top two bands carry opposite Chern number and are separated from the rest of the spectrum, suggesting
 a description in terms of an effective tight binding model.  
We now focus on $\theta < \theta_1$ and directly derive an effective tight binding model for the first two moir\'e bands 
by explicitly constructing a basis of localized Wannier states.
These Wannier states are constructed via a simple procedure which manifestly preserves the symmetries of the twisted homobilayer.
Given the single particle eigenstates $\{\ket{\phi_{n,\textbf{k}}}\}$ of the continuum Hamiltonian~\eqref{eq:Hcont} for each $\textbf{k}$ in the mBZ, we first construct a superposition of the first two $(n=1,2)$ eigenstates,
$\ket{\tilde{\phi}_{n,\textbf{k}}} = \sum_{m=1,2} U^{(\textbf{k})}_{n m} \ket{\phi_{m\textbf{k}}}$ using a $2\times 2$ unitary matrix $U^{(\textbf{k})}_{nm}$,
which maximizes the layer polarization at every $\textbf{k}$:
\begin{equation}
P_{\textbf{k}} = \sum_{n=1,2} (-1)^{n}\braket{\tilde{\phi}_{n,\textbf{k}}|
(\mathcal{P}_{-} - \mathcal{P}_{+})
|\tilde{\phi}_{n,\textbf{k}}} ,
\end{equation}
where $\mathcal{P}_{\pm}$ is the projector on to the top/bottom layer,
so that $\ket{\tilde{\phi}_{1,\textbf{k}}}$ is chosen to mostly consist of states in the top layer, and similarly for $\ket{\tilde{\phi}_{2,\textbf{k}}}$ on the bottom layer. 
This uniquely specifies $\ket{\tilde{\phi}_{n,\textbf{k}}}$ up to a phase, which we choose to be real and positive at the XM ($n=1$) or MX ($n=2$) stacking regions { (Supplementary Note 5~\cite{supp})}.
The Wannier states at moir\'e lattice vector $\textbf{R}$ is then defined  $\ket{W_{\textbf{R}}^n} = \frac{1}{\sqrt{N_k}} \sum_{\textbf{k}} e^{-i \textbf{k}\cdot\textbf{R}} \ket{\tilde{\phi}_{n \textbf{k}}}$.
They are localized about their centers with a root-mean-square distance $a_W\approx 5\text{nm}$, and are also mostly composed of states in one layer:
$\braket{W_{\textbf{R}}^1|\mathcal{P}_+|W_{\textbf{R}}^1}\approx 0.83$ is mostly in the top layer, and vice versa for $\ket{W_{\textbf{R}}^2}$.

\begin{figure}
\includegraphics[width=\columnwidth]{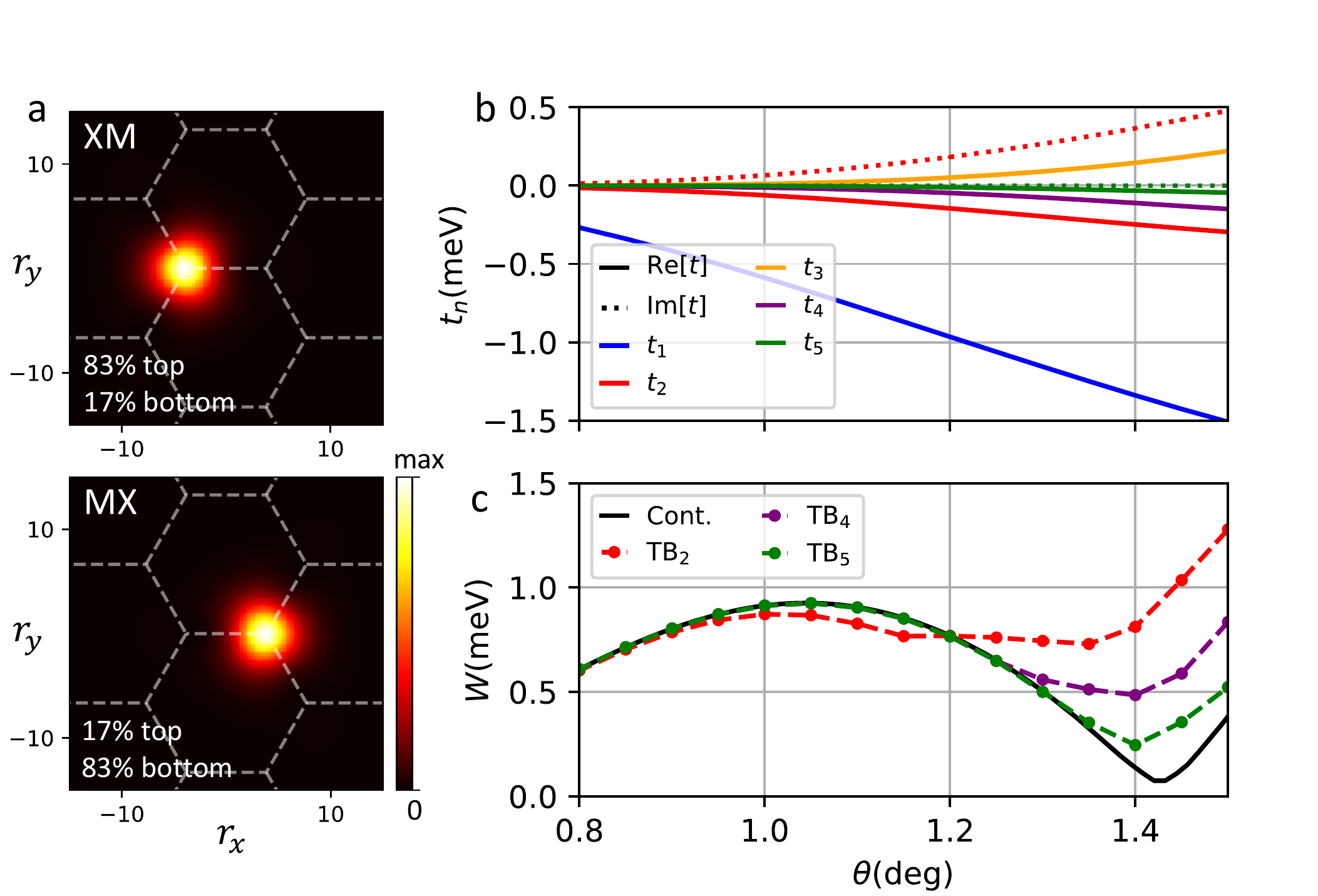}
\caption{
{Wannier functions and tight binding model parameters.}
$a)$ Wannier functions at the magic angle, $b)$ tight binding parameters as a function of $\theta$, and $c)$ the bandwidth of the top band in the effective tight binding model TB$_n$ keeping up to $n$th nearest neighbor hopping terms, compared to that of the continuum model.
}\label{fig:Wannier}
\end{figure}

It is straightforward to obtain the hopping matrix elements of the effective tight binding model in the Wannier basis for the top two bands as a function of $\theta$.
Figure~\ref{fig:Wannier}b shows the $n$th nearest neighbor hopping matrix elements $t_n$ obtained in this way, up to $n=5$.
As anticipated, the effective tight binding model at $\theta<\theta_1$, including the spin/valley degrees of freedom, is a generalized Kane-Mele model with sites centered on the honeycomb lattice formed by MX and XM stacking regions~\cite{Wu2019}.

The tight binding Hamiltonian is found to be
\begin{equation} \label{eq_effKaneMele}
\mathcal{H}_{\mathrm{TB}} = t_1 \sum_{\langle i,j\rangle,\sigma} c^\dagger_{i\sigma} c_{j\sigma} +  |t_2| \sum_{\langle \langle i,j\rangle\rangle,\sigma} e^{i\phi \sigma \nu_{ij}} c^\dagger_{i\sigma} c_{j\sigma}
+\cdots
\end{equation}
where $c^\dagger_{i\sigma}, c_{i\sigma}$ are fermionic creation/annihilation operators, $\sigma=\pm$ is the spin/valley degree of freedom, the sum $\langle i,j\rangle$ ($\langle \langle i,j\rangle\rangle$) is over (next) nearest neighboring sites $i,j$ of the honeycomb lattice, and $\nu_{ij}=\pm 1$ depending on whether the path $i\rightarrow j$ turns right $(+)$ or left $(-)$.
The parameter $t_1$ is real, while $t_2 \equiv |t_2|e^{i\phi}$ is complex, and $\cdots$ contain longer range hopping terms.
We find that $|t_n|$ quickly reduce in magnitude with hopping distance $n$, and only the 2nd neighbor hopping has a significant imaginary component.
In Figure~\ref{fig:Wannier}c, we show the bandwidth of the top band, $W$, in the effective tight binding model TB$_n$ including up to $t_n$ hopping terms, compared to that of the continuum model.
For $\theta\lesssim 1^\circ$, TB$_2$ already captures the band structure very well.
Near the magic angle, higher range hoppings become more important in capturing the flatness of the band.

For the small twist angles $\theta<\theta_1$ considered, 
since the size of the Wannier orbitals are small compared to the moir\'e unit cell, 
the dominant interaction is a simple on-site Hubbard term $\mathcal{H}_U = U \sum_{i} n_{i \uparrow} n_{i\downarrow}$.  
We also estimate $U\sim e^2/(\epsilon a_W) \approx 70 \text{meV}$ at $\theta_m$, using a realistic relative dielectric constant $\epsilon=4$ and $a_W=5\rm{nm}$, which is significantly larger than the tight binding parameters $t_n$. Therefore at such small twist angle, twisted WSe$_2$ homobilayers are in the strong-coupling regime, in contrast with $\theta \sim 4^\circ$--$5^\circ$ where the bandwidth is comparable to the interaction strength \cite{Wang2020,Bi2021}.

In the following, we shall focus on the strongly correlated regime $\theta<\theta_1$ at a filling of $n=1$ holes per moir\'e unit cell, where we expect the flat bands will favor the quantum anomalous Hall (QAH) insulator due to spontaneous spin/valley polarization. The reason is as follows. First, spin/valley polarized states filling the top band of $\mathcal{H}_{\rm TB}$ with $\sigma = \pm$ are exact eigenstates of our interacting model, because the spin-orbit coupling in $\mathcal{H}_{\rm TB}$ conserves the $z$-spin/valley component. Then, these spin/valley polarized states avoid the Hubbard interaction, and also minimize the kinetic energy in the case of a completely flat top band. Minimizing both parts of the Hamiltonian, they necessarily are many body ground states of the model at $n=1$ filling. The complete polarization of $\mathcal{H}_{\rm TB}$ exactly corresponds to Haldane's model for QAH insulator.

\begin{figure}
\includegraphics[width=\columnwidth]{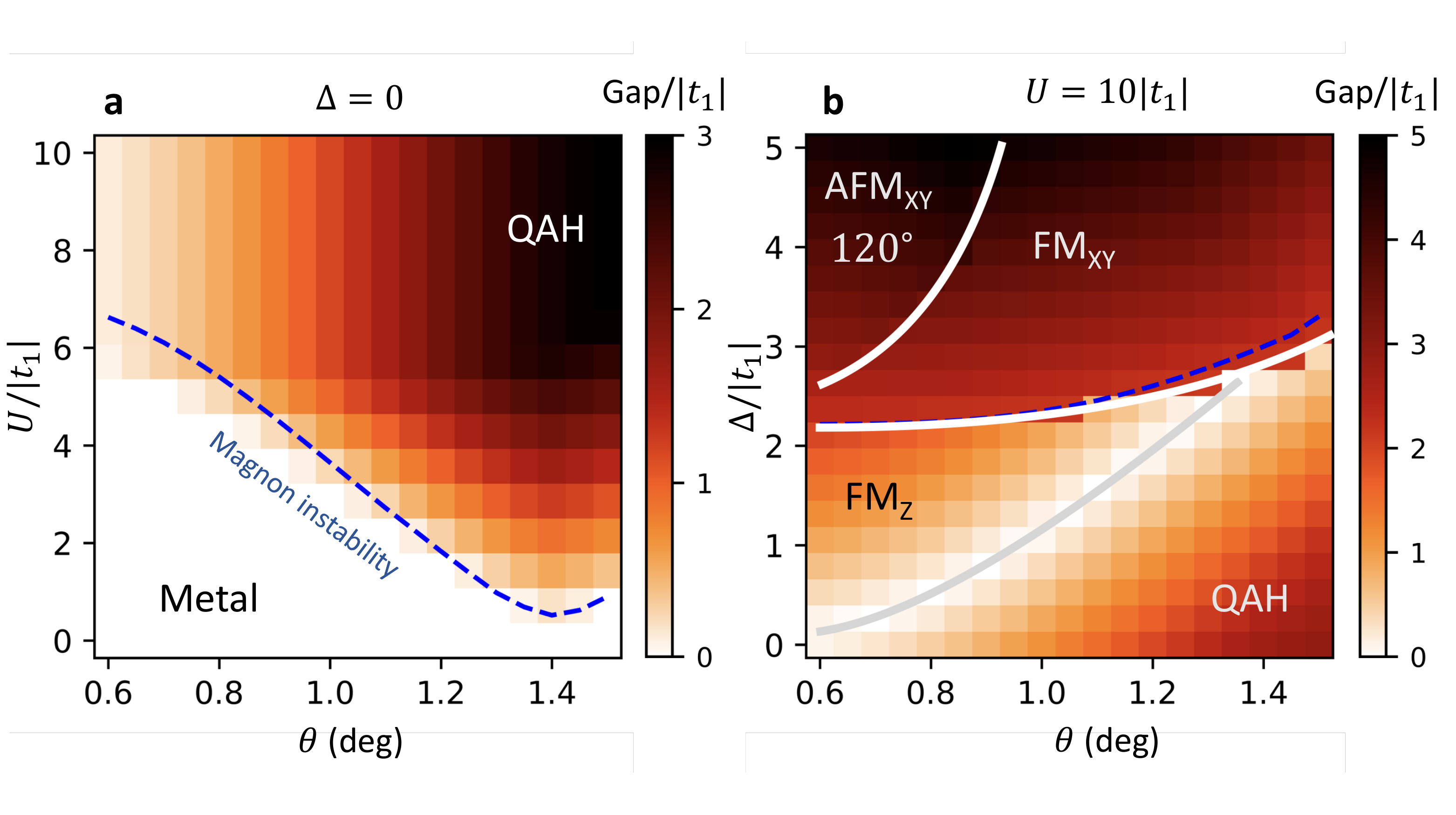}
\caption{
{ Numerical solution of the self-consistent Hartree-Fock approximation.
The phase diagram is} shown as a function of the twist angle and (a) interaction strength at fixed displacement field $\Delta = 0$, or (b) displacement field at fixed interaction strength $U = 10 |t_1|$
, including up to 5th nearest neighbor hopping terms.
The insulating phases denoted by QAH, FM$_z$, 120$^\circ$ AFM$_{xy}$, and FM$_{xy}$ are described in the main text. The dashed blue line shows the boundary of the FM$_z$ phases determined by the first magnon instability.
The colors indicate the charge gap.
Hartree-Fock calculations were done with a $\sqrt{3}\times\sqrt{3}$ unit cell of $6$ atom sites, and a $30\times 30$ grid of $\textbf{k}$ points.
}
\label{fig:AddingU}
\end{figure}

To include corrections coming from the finite bandwidth $W$ of the flat band, we solve our interacting problem within the Hartree-Fock (HF) approximation, where the Hubbard interaction is decoupled as $n_{i \uparrow} n_{i \downarrow} \simeq \langle n_{i \uparrow} \rangle n_{i \downarrow} + \langle n_{i \downarrow} \rangle n_{i \uparrow} - \langle c_{i \uparrow}^\dagger c_{i \downarrow} \rangle   c_{i \downarrow}^\dagger c_{i \uparrow} - \langle c_{i \downarrow}^\dagger c_{i \uparrow} \rangle  c_{i \uparrow}^\dagger c_{i \downarrow}$, up to a constant term, and where the expectation values for the spin and density at each site are determined self-consistently by iteration. Our numerical solutions of the HF equations are shown in Fig.~\ref{fig:AddingU}a as a function of twist angle and interaction strength. As expected, we observe a transition from a metallic state to a ferromagnetic QAH insulator polarized along $z$ when $U$ increases. Within HF, this transition can be understood as follows. The fully polarized states yield a rigid shift of the bands by $\sigma U/2$. When $U$ is larger than the non-interacting bandwidth $W$, a full gap opens and the ferromagnetic state fully fills one of Chern bands of $\mathcal{H}_{\rm TB}$, which leads to a QAH phase. We remark that the appearance of the QAH phase relies on both the non-trivial Chern number as well as the fact that the band is flat and isolated, features which are maximized at the magic angle, 
as illustrated by a dip of the insulating phase above $\theta \simeq 1.4^\circ$.

To precisely locate the transition between QAH insulator and the metallic phase, we compute the magnon excitation spectrum above the fully ferromagnetic state by exact diagonalization (ED) of the interacting Hamiltonian projected on the spin-1 excitation subspace~\cite{EDMagnons}. For large $U$, this spectrum is gapped and the QAH is robust against spin flips. Decreasing $U$ eventually brings one magnon at zero energy, which destabilizes the ferromagnetic states and drives the transition to a metal. As shown in Fig.~\ref{fig:AddingU}a, the ED results almost perfectly agrees with the HF boundaries, putting them on firmer grounds.

For the large values of $U$ relevant to WSe$_2$, the magnons have a large gap, and the lowest excitation corresponds to an interband transition between two bands with same spin. The QAH phase is thus protected by a gap $\varepsilon_{12} \approx 3.7$meV near the magic angle, leading to quantum Hall effect at elevated temperature. 

We also highlight that the QAH may also be observed for larger twist angles, where the first band still carries a non-zero Chern number (Fig.~\ref{fig:BandBerry}), and its bandwidth remain small compared to the estimated $U$ (Fig.~\ref{fig:gaps}).  
Likewise, the second band is topological and quite flat for $\theta\sim 2^\circ\text{--}3^\circ$ (Supplementary Note 3), and therefore QAH may also be observed at a filling of $n=3$.
Twisted TMD bilayers with $\pm K$-valley bands are thus expected to 
be an intrinsically robust platform for interaction-induced QAH phases.

It is interesting and worthwhile to compare the QAH phase in twisted TMD and graphene bilayers. 
Anomalous Hall effect and its quantization have been experimentally observed in 
magic-angle TBG \cite{Sharpe2019,Serlin2020}, where the alignment with hBN substrate
is likely the origin of valley Chern number~\cite{Song2015} and 
both spin and valley degeneracy are lifted due to repulsive interaction in the flat band~\cite{Zhang2019,xie2020nature,liu2021nematic,liu2021theories,Bultinck2020} 
Due to the presence of $SU(2)$-invariant spin degrees of freedom, QAH in TBG is subject to 
the adverse effect of gapless thermal fluctuation, which forbids long-range order at finite temperature in the thermodynamic limit. 
In contrast, spin-valley locking in TMD systems enables robust Ising-type spin/valley order  
that leads to QAH effect at lower temperature.

Another great advantage of twisted TMD bilayer is their high degree of tunability, in particular with respect to applied electric fields~\cite{ElectricallytunableValley,Wang2020,Zhang2020}. 
Due to the layer polarization of the Wannier basis states, the displacement field can be modeled as a sublattice symmetry breaking term $\mathcal{H}_\Delta = \frac{\Delta}{2} \sum_{i} s_i c_{i\sigma}^\dagger c_{i\sigma}$, where $s_i$ is $(-)1$ for $i$ in the $A$ ($B$) sublattice. 
Including this term in our HF treatment, we can investigate which phases should neighbour the QAH ferromagnet in experiments. We present our solutions of the HF equations as a function of twist angle and displacement field in Fig.~\ref{fig:AddingU}b. There, we fix $U=10|t_1|$, a tradeoff between the large $U$ of the homobilayer system and the convergence rate of the HF self-consistent iteration algorithm. 
We find it necessary to consider an enlarged $\sqrt{3} \times \sqrt{3}$ unit cell in order to describe all ordered phases of the model.

At small displacement fields, the topmost moiré band remain relatively flat and our earlier arguments for spin/valley polarization still apply. This is confirmed by our HF solutions for $\Delta \lesssim 2 t_1$ (Fig.~\ref{fig:AddingU}b), which exhibit full spin polarization along the $z$ axis. In this region, a transition nevertheless occurs at $\Delta = 6 \sqrt{3} |t_2| \sin \phi$ (up to $t_{n\geq 3}$ terms), where the single-particle gap between the two moiré bands closes, and their Chern numbers change from $[+1,-1]$ to $[0,0]$. This gap closing line marks the transition between a QAH insulator and a topologically trivial ferromagnet with spin/valley polarization (FM$_z$)~\cite{Haldane1988}. As the displacement field further increases, the bandwidth $W$ also grows, which decreases the magnon gap (see discussion above). The spin/valley polarized phases eventually become unstable when the magnon gap closes, which can be seen with the very good agreement between the phase boundaries determined with HF and ED (Fig.~\ref{fig:AddingU}b).

Beyond this spin-wave instability line, our results show the emergence of two new Mott insulating phases, where holes are mostly localized on the $A$ sublattice, and their spin either form an antiferromagnetic pattern (120$^\circ$ AFM$_{xy}$), or ferromagnetically align in the $xy$ plane (FM$_{xy}$). Their appearance is most easily understood for large displacement fields, where the physics becomes analogous to that of localized moments on the triangular $A$ sublattice. In the regime $t_2 \lesssim t_1 \ll \Delta, U$ relevant for our system, their coupling is described by an effective XXZ model with Dzyaloshinskii-Moriya (DM) interactions
\begin{equation} \label{eq_effectivespinmodel} 
\mathcal{H}_S = \! \sum_{\langle i, j \rangle_B} \! J_\parallel s_i^z s_j^z + J_\perp (s_i^x s_j^x + s_i^y s_j^y) + D \left[ (\textbf{s}_i \times \textbf{s}_j) \cdot \textbf{z} \right] ,
\end{equation}
which is derived in { Supplementary Note 6~\cite{supp}}. The parameters of this effective spin model are given by
\begin{subequations} \label{eq_parametereffectivespin} \begin{eqnarray}
J_\parallel &=& \frac{4 |\tilde{t}|^2}{U} + {\rm Re} \left( \frac{4t_1^2 t_2}{\Delta^2} \right) , \\
J_\perp &=& {\rm Re} \left( \frac{4 \tilde{t}^2}{U} + \frac{4t_1^2 t_2}{\Delta^2} \right) , \\
D &=& {\rm Im} \left( \frac{4 \tilde{t}^2}{U} + \frac{4t_1^2 t_2}{\Delta^2} \right) , \end{eqnarray} \end{subequations}
with $\tilde{t} = t_2 + t_1^2/\Delta$. In Eq.~\ref{eq_parametereffectivespin}, we have separated exchange terms coming from different physical processes. The first ones $\propto \tilde{t}^2/U$ arise from nearest neighbor tunneling on the triangular $A$ sublattice, while the others $\propto t_2 (t_1/\Delta)^2$ originate from loop-exchange on the honeycomb lattice that do not involve any double occupancy.

For twist angles $\theta\lesssim 1^\circ$, $t_2\ll t_1$ and Eq.~\ref{eq_effectivespinmodel} reduces to an antiferromagnetic (AFM) Heisenberg model, where $J_\parallel =J_\perp >0$ are dominated by the nearest neighbor tunneling on the triangular lattice. This simplified triangular lattice description, valid for very small twist angles, has been proposed in earlier studies of Mott insulators in twisted TMDs~\cite{SchradeFu,Pan2020,Zang2021}. It was shown to yield an antiferromagnetic phase that the small residual DM interaction pins in the $xy$ plane. This is the origin of the AFM$_{xy}$ phase observed in Fig.~\ref{fig:AddingU}b. 
We also note that the weak-coupling version of AFM$_{xy}$ phase---an intervalley-coherent $\sqrt{3}\times\sqrt{3}$ density wave \cite{Bi2021}---has been proposed for the correlated insulating state at $n=1$ in twisted bilayer WSe$_2$ at $\theta \sim 4^\circ$--$5^\circ$ \cite{Wang2020}.

For larger twist angles, $t_2$ becomes substantial and we observe that $J_\perp$ becomes negative for the realistic parameter $U\gg \Delta$, 
dominated by a third-order exchange process $\propto t_1^2 t_2$ on the honeycomb lattice without double occupancy. Then,  
the FM$_{xy}$ phase is favored as shown in Fig.~\ref{fig:AddingU}b. 
The competition between AFM$_{xy}$ and FM$_{xy}$ phases can be analyzed by solving Eq.~\ref{eq_effectivespinmodel} for classical spins. 
This approach, detailed in the { Supplementary Note 6~\cite{supp}}, gives a transition between the two phases when $|D| = - \sqrt{3} J_\perp$. For $\Delta=5t_1$, this criterion yields a critical twist angle $\theta = 0.95^\circ$, which roughly agrees with our HF results.
We note that the ferromagnetic phase due to $J_\perp<0$ does not appear in twisted TMDs based on simplified triangular lattice descriptions. 

 Finally, we comment on the effect of nearest neighbor repulsion $V\sum_{\langle i,j\rangle}n_i n_j$ to our HF phase diagram.  
This term favors the layer polarized phases, such as the FM$_{xy}$ and AFM$_{xy}$ which appear at large $|\Delta|$.  
Small $V$ therefore narrows the range in $\Delta$ at which the QAH phase appears.
For large $V$, there is a sharp transition at $\Delta=0$ between layer polarized Mott insulating phases, which can lead to the strong hysteretic behavior of Mott ferroelectricity~\cite{Zheng2020,Zhang2021}.
The long range component of interactions can be controlled by screening from nearby metallic layers.
Multiple recent experiments~\cite{gu2021dipolar,zhang2021correlated} on twisted WSe$_2$ homobilayers in the presence of a nearby WSe$_2$ monolayer report strong screening effects when the monolayer is doped.
This raises the interesting possibility of a screening-induced transition between the QAH and Mott insulating phases.

\section{Discussion}
Our phase diagram demonstrates the high experimental tunability of TMD twisted homobilayers, where the applied displacement field can tune between quantum anomalous Hall phase and Mott insulators involving three types of magnetic orders: FM$_z$, FM$_{xy}$ and AFM$_{xy}$. 
Despite being electrically insulating, the $xy$-ordered Mott insulators support coherent magnon transport  \cite{Bi2021}, 
which can be detected by optical spin injection and spatial-temporal mapping recently developed for TMD bilayers~\cite{Jin2018}.  The experimental feasibility of tuning and distinguishing between topologically different insulators at the same filling 
adds to the attractiveness and desirability of TMD based moir\'e systems.

In parallel to our work on twisted TMD homobilayers, a breakthrough experiment led by Kin Fai Mak and Jie Shan 
discovered unexpectedly a QAH phase with spontaneous spin/valley polarization 
in a TMD heterobilayer MoTe$_2$/WSe$_2$ at the filling of $n=1$ tuned by displacement field~\cite{fai2021,ZhangPNAS}. 
Large-scale DFT calculation and wavefunction analysis reveal two dispersive moir\'e bands forming the Kane-Mele model,  
suggestive of a similar origin of QAH as described here.

Looking forward, the remarkable flat Chern band we found, combined with the uniformity of Berry curvature, suggests that twisted TMD homobilayers near magic angle  may be an ideal setting for observing a fractional quantum anomalous Hall state at zero magnetic field.

{
\section{Data availability}
The data needed to evaluate the conclusions in the paper are present in the paper and the Supplementary Material. 
The full dataset generated during this study, including relaxed lattice structure and band structure obtained from DFT, tight binding model parameters, and self-consistent HF solutions, have been deposited in the Zenodo database~\cite{data}.
Additional data related to this paper is available from the corresponding author upon reasonable request.
}


\section{Acknowledgment}
We thank Kin Fai Mak, Jie Shan, Tingxin Li and Shengwei Jiang  for ongoing collaborations on 
MoTe$_2$/WSe$_2$,  Bi Zhen and Constantin Schrade for previous collaborations on related topics. 
We thank Pablo Jarillo-Herrero, Kenji Yasuda, Cory Dean, Abhay Pasupathy, Qianhui Shi, Augusto Ghiotto and En-Min Shih for helpful discussions.   

This work is primarily supported by DOE Office of Basic Energy Sciences, Division of Materials Sciences and Engineering under Award DE-SC0020149 (band structure calculation), DE-SC0018945 (theoretical modeling) and 
Simons Investigator award from the Simons Foundation (numerical analysis). 
LF is partly supported by the David and Lucile Packard Foundation. 

{
\section{Author Contributions}

T.D., V.C., Y.Z. and L.F. performed research, analyzed data, and wrote the manuscript.

\section{Competing interests}
The authors declare no competing interests.
}

\end{document}


\begin{center}
\textbf{\large Supplementary Information for ``Magic in twisted transition metal dichalcogenide bilayers"} \\[10pt] 
Trithep  Devakul, Valentin  Cr\'epel, Yang  Zhang, and Liang  Fu \\
\textit{Department of Physics, Massachusetts Institute of Technology, Cambridge, Massachusetts 02139, USA}
\end{center}
\vspace{20pt}
\twocolumngrid

\section{Supplementary Methods}
\subsection{Supplementary Note 1}
\emph{Density functional theory calculation and continuum model} ---
We study TMD homobilayers with a small twist angle starting from AA stacking, where every metal (M) or chalcogen (X) atom  on  the top layer is aligned with the same type of atom on the bottom layer (AB stacking can be viewed as a $180^{\circ}$ rotation of top layer). 
Within a local region of a twisted bilayer, the atom configuration is identical to that of an untwisted bilayer, where one layer is laterally shifted relative to the other layer by a corresponding displacement vector ${\bm d}_0$. For this reason, the moir\'e band structures of twisted TMD bilayers can be constructed from a family of untwisted bilayers at various ${\bm d}_0$, all having $1\times 1$ unit cell. Our analysis thus starts from untwisted bilayers.

In particular, ${\bm d}_0=0, \left(-{\bm a}_{1}+{\bm a}_{2}\right) /3, \left({\bm a}_{1}+{\bm a}_{2}\right) /3$, where ${\bm a}_{1,2}$ is the primitive lattice vector  for untwisted bilayers, correspond to three high-symmetry
stacking configurations of untwisted TMD bilayers, which we refer to as MM, XM, MX. In MM (MX) stacking, the M atom on the top layer is locally aligned with the M (X) atom on the bottom layer, likewise for XM. The bilayer structure in these stacking configurations is invariant under three-fold rotation around the $z$ axis.

\begin{figure}
\includegraphics[width=1.0\columnwidth]{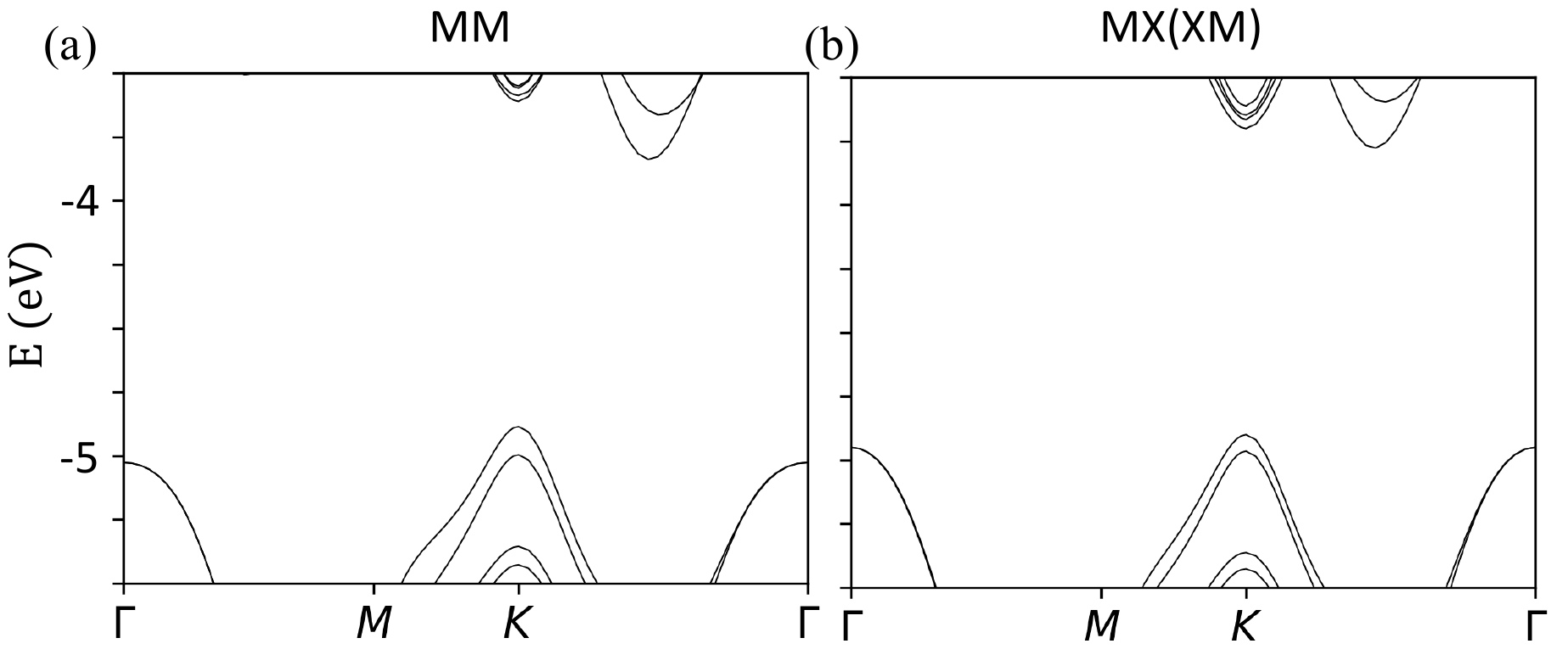}
\caption{Band structure with reference energy $E_0$ set as the absolute vacuum level in (a) MM stacking structure; (b) MX (XM) stacking structure. }
\label{figS1}
\end{figure}

Density functional calculations are performed using generalized gradient approximation with SCAN+rVV10 Van der Waals density functional, as implemented in the Vienna Ab initio Simulation Package. Pseudopotentials are used to describe the electron-ion interactions. We first construct the zero-twisted angle WSe2/WSe2 bilayer at MM and MX lateral configurations with vacuum spacing larger than 20\AA{} to avoid artificial interaction between the periodic images along the z direction. The structure relaxation is performed with force on each atom less than 0.01 eV/A. We use $12\times 12 \times 1$ for structure relaxation and self-consistent calculation. The more accurate SCAN+rVV10 van der Waals density functional gives the relaxed layer distances as 6.6\AA{} and 7.15\AA{} for MX and MM stacking structures, respectively. By calculating the work function from electrostatic energy of converged charge density, we plot in Fig. \ref{figS1} the band structure of MM and MX-stacked bilayers, with reference energy $E=0$ chosen to be the absolute vacuum level. 

\begin{figure}
\includegraphics[width=0.9\columnwidth]{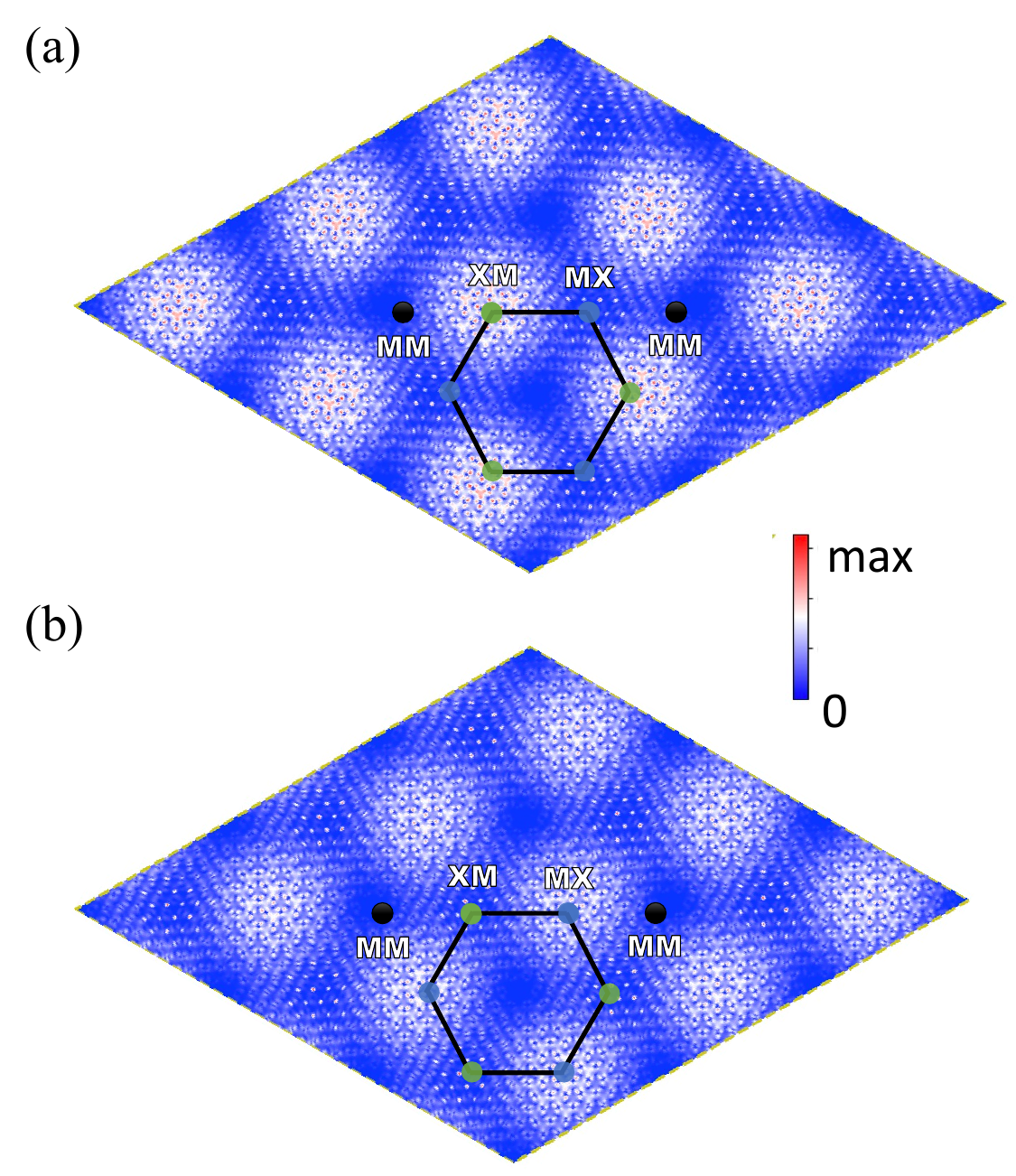}
\caption{$\gamma$ point Kohn-Sham Wavefunction of the topmost moir\'e valence band projected at (a) top layer, (b) bottom layer. }
\label{figS2}
\end{figure}

To compare with moir\'e band structure, we further construct the commensurate structures with twist angle $5.08^\circ$ with 762 atoms, and perform large scale DFT calculation to relax the moir\'e superlattice. The spacial dependent layer distance profile has been discussed in the main text, and fits nicely with the untwisted MM and MX stacking structures. The moir\'e band structure with spin orbital coupling agrees remarkably well with the continuum model derived from untwisted calculations, and the effective mass is chosen as 0.38$m_e$, close to the experimental value 0.4$\sim$0.45$m_e$. To reveal the real space localization of moir\'e flat band, the $\gamma$ point Kohn-Sham wavefunction for the top-most moir\'e valence band is calculated and projected to bottom and top layer, respectively. As shown in Fig. \ref{figS2}, we find that the charge are localized at MX and XM region, consistent with previous Wannier function.

\begin{figure}
\includegraphics[width=0.9\columnwidth]{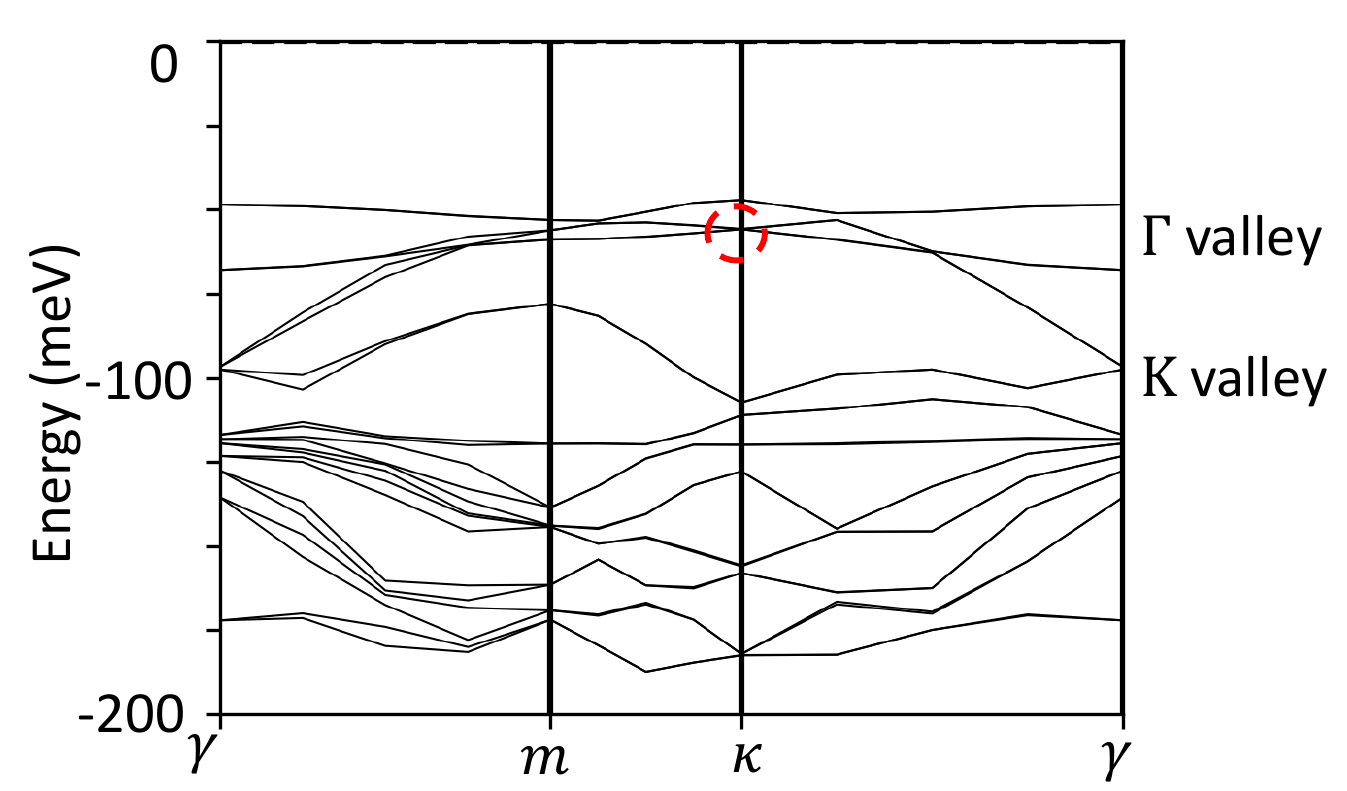}
\caption{Moir\'e band structure of fully relaxed AA stacked homobilayer MoTe$_2$ at $\theta=5.08^\circ$. $\Gamma$ valley moire bands and $K$ valley moire bands are very close in energy, and we circle the Dirac points from $\Gamma$ valley moire bands.}
\label{figS-MoTe2}
\end{figure}

For MoTe$_2$, the $\Gamma$ pockets appears to be the valence band maximum at MX stacking structure when using lattice constant 3.52\AA{} from bulk structures (The relaxed layer distance is 6.95\AA{}, close to the bulk layer distance 6.98\AA{} in 2H structures). Experimentally, AB stacked bilayer MoTe$_2$ has been shown to be a indirect band gap semiconductor from optical measurements \cite{lezama2015indirect,ruppert2014optical}. We thus believe $\Gamma$ pockets would enter in the moir\'e valence bands for twisted homobilayer MoTe$_2$ even for AA stacking. At $\theta=5.08^\circ$, we calculate the moir\'e band structure of fully relaxed AA stacked homobilayer MoTe$_2$. As shown in Fig. \ref{figS-MoTe2}, the narrow Dirac like $\Gamma$ valley moire bands intersect with topmost dispersive $K$ valley moire bands. In transport measurements under zero displacement field, these two type of charge carriers would both enter, causing further complications.

\subsection{Supplementary Note 2}
\emph{Moir\'e topology criteria} ---
We determine the Chern number of top two moir\'e valence bands from their $C_{3z}$ eigenvalues. Computationally, the Kohn-Sham wavefunctions in plane-wave basis $$
\left|\Psi_{n \boldsymbol{k}}\right\rangle=\sum_{\boldsymbol{G}} C_{n \boldsymbol{k}}(\mathbf{G})|\boldsymbol{k}+\boldsymbol{G}\rangle
$$
are extracted at three $C_{3z}$ invariant momentum $\gamma$, $\kappa$, and $\kappa'$ with energy cutoff as $E_{cut}=100$ eV and $\hbar^{2}(\boldsymbol{k}+\boldsymbol{G})^{2} / 2 m_{e}<E_{\mathrm{cut}} $. And we compute their $C_{3z}$ eigenvalue as:
\begin{equation}
\left\langle\Psi_{n k}|C_{3z}| \Psi_{n k}\right\rangle=\sum_{\tau \tau^{\prime}} S_{\sigma \sigma^{\prime}}(C_{3z})\left\langle\Psi_{n k}^{\sigma}|C_{3z}| \Psi_{n k}^{\sigma^{\prime}}\right\rangle
\end{equation}
where $\sigma,\sigma'=\uparrow$ or $\downarrow$, $\left\langle\Psi_{n k}^{\sigma}|C_{3z}| \Psi_{n k}^{\sigma^{\prime}}\right\rangle$ calculated from the transformation relation of plane-waves, and $S_{\sigma \sigma^{\prime}}(C_{3z})$ is the $C_{3z}$ spin rotation matrix.

As there is a valley time reversal symmetry within the moire superlattice, the energy bands between $\gamma$, $\kappa$ and $\kappa'$ are enforced to be double degenerate from $K$ and $-K$ valley. Therefore, to determine the valley index of moir\'e bands, we add a small gating field $D=0.005$ V/nm to introduce a less than 2 meV splitting at $\kappa$ and $\kappa'$. The $K$ and $-K$ indexes are then diagnosed by spin resolved wavefunction.
At $\gamma$ point where bands are still double degenerate with gating field, we extracted the $C_{3z}$ eigenvalues and valley indexes from finite momentum perturbation along $\gamma$ to $\kappa$ direction.

\subsection{Supplementary Note 3}
\begin{figure*}
  \includegraphics[width=\textwidth]{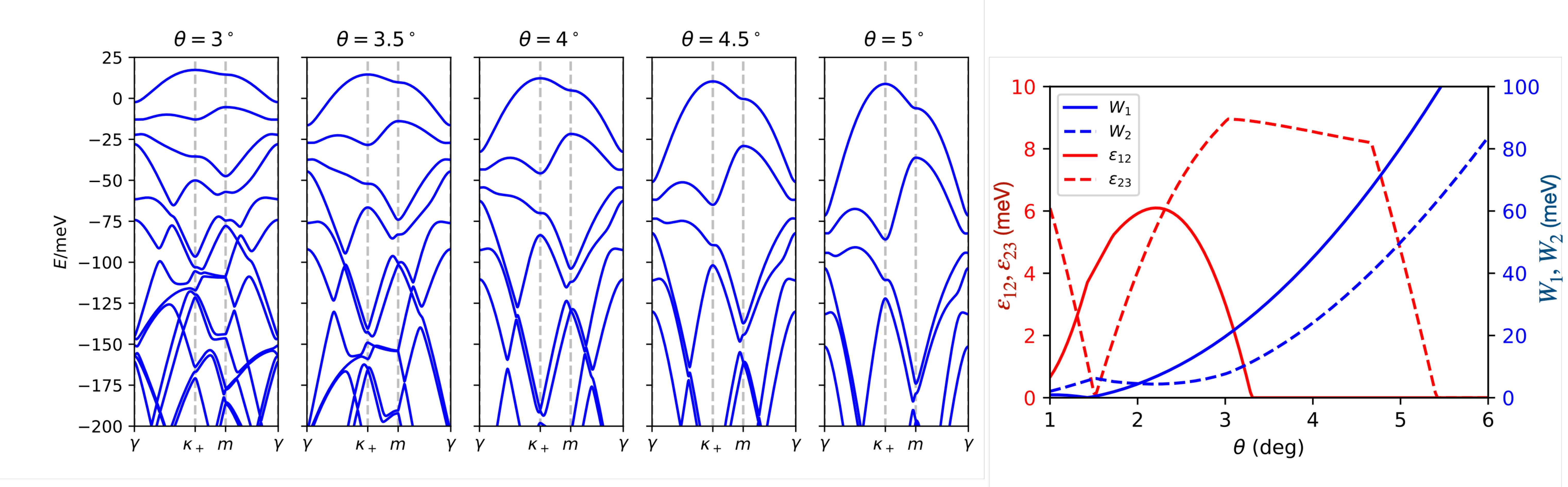}
  \caption{(left) Band structure of the continuum model at twist angles $\theta=3^\circ$--$5^\circ$, larger than shown in the main text.
  (right) The bandwidth $W_1,W_2$ of the first and second bands and the indirect energy gaps between the first two bands, $\varepsilon_{12}$, and between the second and third bands $\varepsilon_{23}$.
  }\label{fig:SMBands}
\end{figure*}

\emph{Continuum model at higher angles} ---
Figure~\ref{fig:SMBands} show the band structure for the continuum model at larger twist angles than discussed in the main text, as well as the band gaps and bandwidth.
The bandwidth of the first band $W_1$ increases rapidly for $\theta>\theta_1\approx1.5^\circ$, reaching $W\sim 100\text{meV}$ for $\theta\sim 5^\circ$--$6^\circ$.
The bandwidth of the second band $W_2$ has a minima at $\theta\approx2.2^\circ$, and remains quite small throughout the angle range $\theta\sim 2^\circ$--$3^\circ$, with a bandwidth $W_2\approx 5\text{meV}$.
Although $\varepsilon_{12}$ vanishes at $\theta_2\approx3.3^\circ$, $\varepsilon_{23}$ remains positive until $\theta\approx 5.4^\circ$.  
For $1.5^\circ < \theta < 5.4^\circ$, the top two bands are thus separated from the rest of the spectrum and form a combined $\mathcal{C}=2$ manifold.

\subsection{Supplementary Note 4}
\emph{Analytic magic angle} ---
In this section, we analytically estimate the magic angle by consider the effective mass of the top band near $\gamma$.

At $k=\gamma$, when $w=V=0$, the highest energy states are sixfold degenerate plane wave states at energy $e_0=-\frac{|\gamma-\kappa_\pm|^2}{2m^*}=-\frac{8\pi^2\theta^2}{9 m^* a_0^2}$, which is separated from the next set of eigenstates at $e_1=-\frac{32\pi^2\theta^2}{9 m^* a_0^2}$.
Incorporating the effect of $w$ and $V$ will introduce coupling between these states at $e_0$, as well as with those at $e_1$ and beyond.
For the purposes obtaining an analytic estimate for the magic angle, we include only this set of 6 highest energy states. 
This approximation is valid for $w,V\ll e_0-e_1$:
as a rough estimate of the validity, $e_0-e_1\approx 130\text{meV}$ at $\theta=1.5^\circ$ is an order of magnitude greater than $(V,w)=(9,18)\text{meV}$ for WSe$_2$. 

These six states consist of the plane waves at $\bm{k}_{j}$ for $j=1\dots 6$, where 
\begin{equation}
\bm{k}_j=
\begin{cases}
     \bm{\kappa}_+ + (\mathcal{R}_{\pi/3})^{j-1}\bm{\delta} & \text{if $j$ odd}\\
     \bm{\kappa}_- + (\mathcal{R}_{\pi/3})^{j-1}\bm{\delta} & \text{if $j$ even}
\end{cases}
\end{equation}
and the $j$ odd (even) states are on the top (bottom) layer, which we label by $\ket{j}$.
Here, $\bm{\delta}=[2\pi\theta/(3 a_0)](\sqrt{3},-1)$ and $R_{\theta}$ is a $\theta$ counter-clockwise rotation matrix.
The continuum Hamiltonian acts on these states as
\begin{equation}
\begin{split}
\mathcal{H}_{\uparrow} \ket{j} = e_0 \ket{j}& + V \left( e^{i\psi} \ket{j+2}+e^{-i\psi} \ket{j-2}\right)
\\
&+ w\left(\ket{j+1}+\ket{j-1}\right) + \cdots
\end{split}
\end{equation}
where $\ket{j\pm 6}\equiv\ket{j}$, and $\cdots$ contain states outside of $\{\ket{j}\}$, which we have ignored due to the large energy gap $e_1-e_0$.
Utilizing the emergent $C_6$ symmetry which sends $\ket{j}\rightarrow\ket{j+1}$,
 $\mathcal{H}_\uparrow$ is easily diagonalized (in the restricted subspace) by eigenstates
 \begin{equation}
 \ket{n} = \frac{1}{\sqrt{6}}\sum_{j=1}^6 e^{i n\pi j/ 3} \ket{j}
 \end{equation}
 with eigenvalue $e_0 + \mathcal{E}_n$, where
 \begin{equation}
 \mathcal{E}_n = 2 w \cos(\pi n / 3) + 2 V \cos(2 \pi n / 3 - \psi).
 \end{equation}
 In the absence of any accidental degeneracy, there is a unique highest energy eigenstate $\ket{n_0}$.
 
 Next, in order to estimate when $\bm{\gamma}$ transitions from band maxima to minima, we consider a small momentum shift about $\bm{\gamma}$.  
 The $C_3$ symmetry of the microscopic model restricts the lowest order momentum dependence to be simply $\frac{1}{2 m_\gamma}|\bm{k}-\bm{\gamma}|^2$ for some effective mass $m_\gamma$.
 Thus, we consider a momentum shift $\epsilon\hat{x}$ away from $\bm{\gamma}$ and work perturbatively in $\epsilon$.
Expanding the energy $E = E^{(0)} + \epsilon^2 E^{(2)}$, we have that $E^{(2)}=1/(2 m_\gamma)$, and the magic angle $\tilde{\theta}_m$ is where $E^{(2)}$ changes sign.

We expand $\mathcal{H}_{\uparrow}$ as $\mathcal{H}_{\uparrow} = H_0 + \epsilon H_1 + \epsilon^2 H_2$, which act on $\ket{j}$ as 
\begin{equation}
H_1 \ket{j} = -\frac{\hat{x}\cdot (\mathcal{R}_{\pi/3})^{j-1} \bm{\delta}}{m^*}\ket{j}
\end{equation}
and $H_2 = -\frac{1}{2 m^*}$.
We have
\begin{equation}
\hat{x}\cdot (\mathcal{R}_{\pi/3})^{j-1} \bm{\delta} = \frac{4\pi\theta}{3 a_0}\sin(j \pi/3) 
\end{equation}
hence, acting on $\ket{n}$,
\begin{equation}
H_1\ket{n} = \frac{2\pi\theta}{3 m^* a_0}\left(\ket{n+1}+\ket{n-1}\right).
\end{equation}
Applying standard perturbation theory in the $\ket{n}$ basis, the second order correction to the energy is
\begin{equation}
\begin{split}
E^{(2)} =& \braket{n_0|H_2|n_0} + \sum_{n\neq n_0}\frac{\braket{n_0|H_1|n}\braket{n| H_1 |n_0}}{\mathcal{E}_{n_0} - \mathcal{E}_{n}}\\
=& -\frac{1}{2 m^*} + \frac{4\pi^2\theta^2}{9 m^{*2}a_0^2}\left(\frac{1}{\mathcal{E}_{n_0}-\mathcal{E}_{n_0+1}}+ \frac{1}{\mathcal{E}_{n_0}-\mathcal{E}_{n_0-1}}\right)
\end{split}
\end{equation}
Solving for where $E^{(2)} = 0$ gives the magic angle condition from the main text,
\begin{equation}
\tilde{\theta}_m^{-2} = \frac{8 \pi^2}{9 m^* a_0^2}\left(\frac{1}{\mathcal{E}_{n_0}-\mathcal{E}_{n_0+1}}+ \frac{1}{\mathcal{E}_{n_0}-\mathcal{E}_{n_0-1}}\right)
\end{equation}

\subsection{Supplementary Note 5}\label{app_wanniers}
\emph{Localized Wannier functions and tight binding model} ---
In this section, we provide additional details on the construction of the Wannier states shown in the main text.
We numerically diagonalize the Hamiltonian $\mathcal{H}_\uparrow$ on an $N_k=30\times 30$ discretization of the moir\'e Brillouin zone, corresponding to a system of $30\times 30$ moir\'e unit cells.
For each $\bm{k}$, the eigenstates $\{\ket{\phi_{n,\bm{k}}}\}_{n}$ is obtained in the plane wave basis, expressed as $\ket{\phi_{n\bm{k}}} = \sum_{\bm{g},l} c^{(n\bm{k})}_{\bm{g}l}\ket{\bm{k}+\bm{g},l}$, where $\bm{g}$ are reciprocal moir\'e lattice vectors, $l=\pm$ is the layer, and $c^{(n\bm{k})}_{\bm{g} l}$ are complex numbers obtained from exact diagonalization.
The diagonalization is done by truncating to some maximum $|\bm{k}+\bm{g}|$ which is sufficiently far from the valence band top at $\bm{\kappa}_\pm$ so as to have negligible effect on the first few moir\'e bands.

Constructing Wannier states for the top two bands involves finding the appropriate $2\times 2$ unitary matrices $U^{(\bm{k})}_{nm}$ at each $\bm{k}$.
For convenience, we shall suppress all $\bm{k}$ labels.
We parameterize the unitary matrix as
\begin{equation}
U = 
\begin{pmatrix}
e^{i\psi_1}\cos(\chi/2) & e^{i\psi_1+i\delta} \sin(\chi/2) \\
e^{i\psi_2-i\delta}\sin(\chi/2) & -e^{i\psi_2} \cos(\chi/2)
\end{pmatrix},
\end{equation}
where $\chi\in[0,\pi]$, and $\delta,\psi_1,\psi_2\in[0,2\pi]$.
As stated in the main text, $U$ is chosen to maximize the layer polarization of $\ket{\tilde{\phi}_n} = \sum_{m} U_{nm}\ket{\phi_m}$,
\begin{equation}
P = \sum_{n=1,2} (-1)^{n}\braket{\tilde{\phi}_{n}|
(\mathcal{P}_{-} - \mathcal{P}_{+})
|\tilde{\phi}_{n}} ,
\end{equation}
where $\mathcal{P}_{\pm}$ projects to the $\pm$ layer.  
Using $\mathcal{P}_- = 1 - \mathcal{P}_+$, this can be simplified to
\begin{equation}
\begin{split}
P &= -2\sum_{n=1,2}(-1)^n \braket{\tilde{\phi}_n |\mathcal{P}_+|\tilde{\phi_n}}\\
&= -2 \sum_{n,m,l} (-1)^n \braket{\phi_m| U^\dagger_{mn} \mathcal{P}_+ U_{nl} |\phi_l}\\
&= 2 \left[\cos(\chi)(P_{11}-P_{22}) + \sin(\chi)(e^{i\delta}P_{12} + e^{-i\delta} P_{21})\right]
\end{split}
\end{equation}
where $P_{nm} = \braket{\phi_n|\mathcal{P}_+|\phi_m}=P_{mn}^*$.

It is straightforward to maximize $P$ with respect to $\chi$ and $\delta$.
First, setting $\partial_{\delta} P = 0$ gives $\delta = \arg P_{21} + s \pi$, where $s\in\{0,1\}$.
Then, setting $\partial_\chi P = 0$ gives
\begin{equation}
\chi_s = \arctan\left(\frac{2(-1)^s |P_{12}|}{P_{11}-P_{22}}\right)\in[0,\pi]
\end{equation}
(notice the range of $\arctan$ is taken to be $[0,\pi]$).
The two solutions for $s=0,1$ correspond to the maximum and minimum of $P$.  
To find the maximum, we simply compare the value of $P$ for these two solutions and take the larger one.

To determine the phase factor $\psi_{1,2}$, we take the real space component of the wavefunction at the MX/XM stacking regions.
Define $\bm{r}_1 = (a_0/2\theta)[-1/\sqrt{3},1]$ and $\bm{r}_2 = (a_0/2\theta)[1/\sqrt{3},1]$ to be the locations of the XM and MX stacking region.
We then take the phase factors $\psi_{1,2}$ such that $\braket{\bm{r}_n,l_n|\tilde{\phi}_1}$ is real and positive, where $l_{1(2)}=+(-)$ (here, $\ket{\bm{r},l}$ is the usual real space state at $\bm{r}$ on layer $l$, satisfying $\braket{\bm{r},l|\bm{k}+\bm{g},l^\prime} = e^{i (\bm{k}+\bm{g})\cdot \bm{r}}\delta_{l l^\prime}$).
This fixes the phase factors $\psi_n = -\arg \braket{\bm{r}_n,l_n|\tilde{\phi}_n}$.

Finally, the Wannier functions localized at $\bm{R}+\bm{r}_n$, for a moir\'e lattice vector $\bm{R}$, are defined
\begin{equation}
\ket{W^{n}_{\bm{R}}} = \frac{1}{\sqrt{N_k}}\sum_{\bm{k}} e^{-i \bm{k} \cdot \bm{R}} \ket{\tilde{\phi}_{n,\bm{k}}}.
\end{equation}
The Wannier centers $\bm{R}+\bm{r}_n$ form a moir\'e honeycomb lattice on the XM and MX stacking regions.

\begin{figure}
\includegraphics[width=0.5\columnwidth]{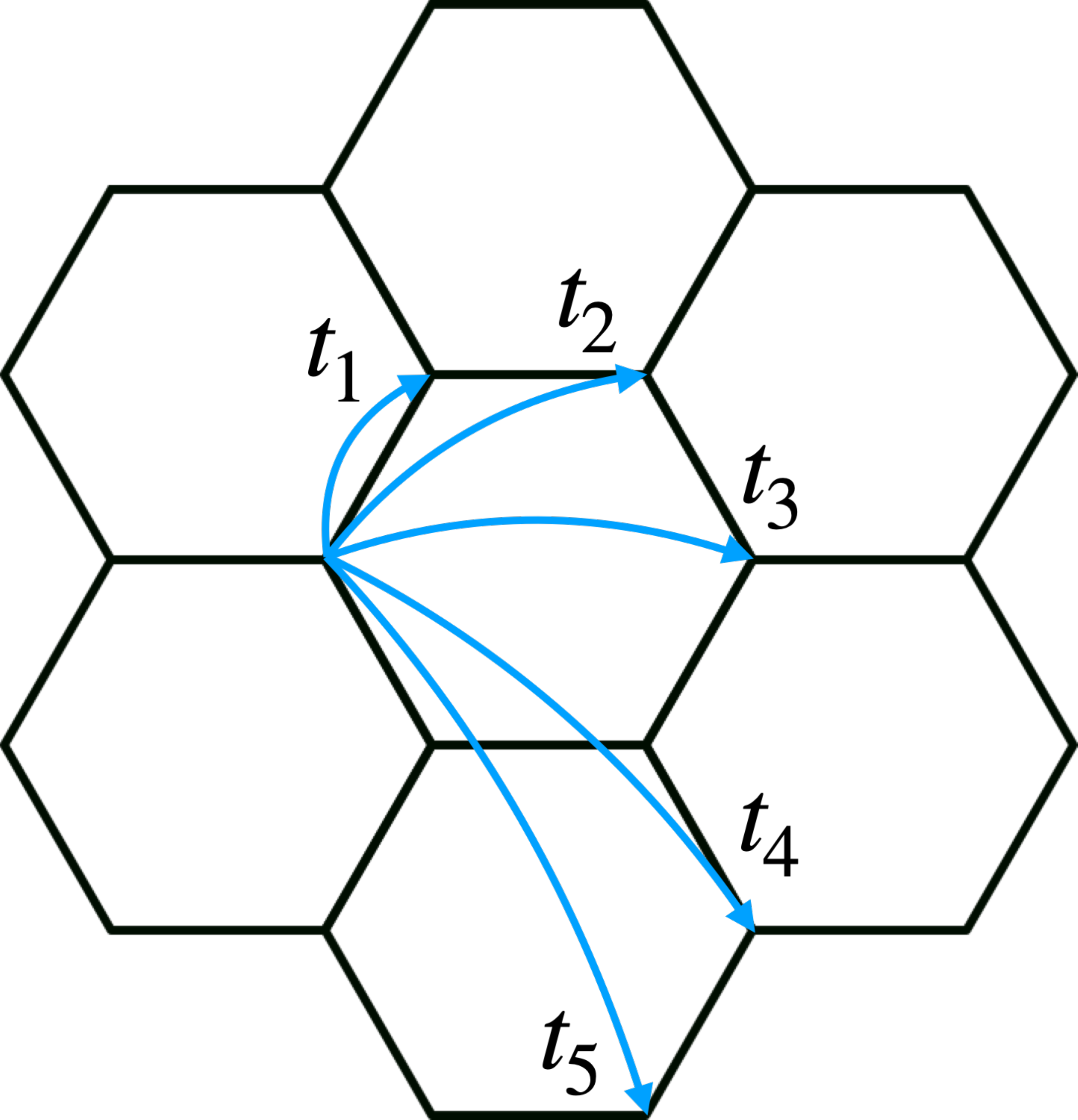}
\caption{Illustration of the hopping matrix elements $t_n$ in the effective tight binding model for spin-$\uparrow$ electrons.}
\label{fig:t5hopping}
\end{figure}

The hopping matrix elements between Wannier states on the site $\bm{R}+\bm{r}_n$ and $\bm{R}^\prime + \bm{r}_{n^\prime}$ can be obtained from the continuum Hamiltonian and $U$ matrices.
We have
\begin{equation}
\braket{W_{\bm{R}^\prime}^{n^\prime}|\mathcal{H}_\uparrow|W_{\bm{R}}^n}
=\frac{1}{N_k}\sum_{\bm{k}} e^{i\bm{k}\cdot(\bm{R}^\prime-\bm{R})} \sum_{m} U^{(\bm{k})\dagger}_{m n^\prime} E_m(\bm{k}) U_{n m}^{(\bm{k})}
\end{equation}
where $E_m(\bm{k})$ is the energy of $\ket{\phi_{m\bm{k}}}$ under $\mathcal{H}_{\uparrow}$.
Using this method, we compute the hopping matrix elements between $n$th nearest neighbor sites up to $n=5$ as illustrated in Fig~\ref{fig:t5hopping}, and shown in the main text.
We find that $t_n$ are mostly real with the exception of $t_2$, which we define (see Fig~\ref{fig:t5hopping}) to be the 2nd nearest neighbor hopping term along right-turning path.  
From the symmetries of the model, it follows that the hopping term along a left-turning path is $t_2^*$.
The $-K$ valley produces a similar tight binding model of spin-$\downarrow$ electrons related to this one by time reversal symmetry.
As a final remark, we note that the Wannier states obtained through this method are not maximally localized, thus our tight binding model has the potential to be further improved.

\subsection{Supplementary Note 6}
\emph{Effective Spin Hamiltonian} ---
The derivation of an effective spin Hamiltonian from a perturbative expansion of Eq.~6 in the limit $|t_1|,|t_2| \ll \Delta, U$ is rather complicated when no clear scale separation exists between $\Delta$ and $U$. Fortunately in our model, the interaction scale $U$ largely dominates over the tunneling terms of $\mathcal{H}_{\rm TB}$, and all experimentally relevant applied displacement fields $\Delta$. As a consequence, we may first eliminate all doubly occupied configurations, and then use $t_1 \ll \Delta$ to obtain the effective spin Hamiltonian given in the main text. 

\subsubsection{Strong coupling expansion}

Let us first fix some notations. We write the interacting Kane-Mele model as 
\begin{equation}
\mathcal{H}_{\rm tot} = \mathcal{H}_U + \mathcal{T}_{-} + \mathcal{T}_{0} + \mathcal{T}_{+} ,
\end{equation}
where $\mathcal{T}_{u}$ gathers all single-particle terms which change the number of doubly occupied by $u$. To get their explicit expression, we gather in $T_{\delta,u}^{(1)}$ the nearest neighbor tunneling processes changing the number of occupied $A$ site by $\delta$ and the number of doubly occupied sites by $u$, where $\delta = \pm$ and $u=0,\pm$. A similar definition $T_{0,u}^{(2)}$ applies for next-nearest neighbor hoppings. These definition yield
\begin{subequations} \begin{eqnarray}
\mathcal{T}_0 &=& \mathcal{H}_\Delta + T_{0,0}^{(2)} + T_{+,0}^{(1)} + T_{-,0}^{(1)} , \\
\mathcal{T}_\pm &=& T_{0,\pm}^{(2)} + T_{+,\pm}^{(1)} + T_{-,\pm}^{(1)} .
\end{eqnarray} \end{subequations}

Elimination of doubly occupied configurations is achieved by standard perturbation theory techniques, see for instance Ref.~\cite{stein1997flow}, and gives 
\begin{equation} \label{appeq_EliminateU} \begin{split}
\mathcal{H}_{\rm tot}^{U} & = \mathcal{T}_{0} - \frac{1}{U} \mathcal{T}_{-} \mathcal{T}_{+} \\
& = \mathcal{H}_\Delta + \tilde{\mathcal{T}}_{-2} + \tilde{\mathcal{T}}_{-1} + \tilde{\mathcal{T}}_{0} + \tilde{\mathcal{T}}_{+1}  + \tilde{\mathcal{T}}_{+2} ,
\end{split} \end{equation}
that we have recast in a form more amenable to perturbative expansion in $\Delta$ by introducing 
\begin{align} 
\tilde{\mathcal{T}}_{0} &=  T_{0,0}^{(2)} - \frac{1}{U} \left[ T_{0,-}^{(2)} T_{0,+}^{(2)} + T_{+,-}^{(1)} T_{-,+}^{(1)}  + T_{-,-}^{(1)} T_{+,+}^{(1)} \right] , \notag \\
\tilde{\mathcal{T}}_{\pm 1} &= T_{\pm,0}^{(1)} - \frac{1}{U} \left[ T_{0,-}^{(2)} T_{\pm,+}^{(1)} + T_{\pm,-}^{(1)} T_{0,+}^{(2)} \right] , \notag \\
\tilde{\mathcal{T}}_{\pm 2} &= - \frac{1}{U} T_{\pm,-}^{(1)} T_{\pm,+}^{(1)}  .
\end{align} 

Similar to the previous case, $\tilde{\mathcal{T}}_\delta$ changes the number of occupied $A$ sites by $\delta$. When their potential energy is large $t_1 \ll \Delta$, these $A$-sites can be eliminated in a similar way as before, see Ref.~\cite{knetter2000perturbation} for details. This leads to the following effective Hamiltonian in the physical subspace with no double occupations and no occupied $A$ sites:
\begin{align} 
& \mathcal{H}_{\rm tot}^{U \Delta} = \tilde{\mathcal{T}}_{0} - \frac{1}{\Delta} \tilde{\mathcal{T}}_{-1} \tilde{\mathcal{T}}_{+1} - \frac{1}{2 \Delta} \tilde{\mathcal{T}}_{-2} \tilde{\mathcal{T}}_{+2} \notag \\ 
& + \frac{1}{8\Delta^2}\left[ 8 \tilde{\mathcal{T}}_{-1} \tilde{\mathcal{T}}_{0} \tilde{\mathcal{T}}_{+1} +4 \left( \tilde{\mathcal{T}}_{-1} \tilde{\mathcal{T}}_{-1} \tilde{\mathcal{T}}_{+2} + \tilde{\mathcal{T}}_{-2} \tilde{\mathcal{T}}_{+1} \tilde{\mathcal{T}}_{+1} \right) \right. \notag \\
& \qquad\quad  - 4 \left( \tilde{\mathcal{T}}_{-1} \tilde{\mathcal{T}}_{+1} \tilde{\mathcal{T}}_{0} + \tilde{\mathcal{T}}_{0} \tilde{\mathcal{T}}_{-1} \tilde{\mathcal{T}}_{+1} \right) + 2 \tilde{\mathcal{T}}_{-2} \tilde{\mathcal{T}}_{0} \tilde{\mathcal{T}}_{+2} \notag \\
& \qquad\quad \left. - \left( \tilde{\mathcal{T}}_{-2} \tilde{\mathcal{T}}_{+2} \tilde{\mathcal{T}}_{0} + \tilde{\mathcal{T}}_{0} \tilde{\mathcal{T}}_{-2} \tilde{\mathcal{T}}_{+2} \right) \right] + \cdots
\end{align}
This expression can be greatly simplified by noticing that the direct action of $T_{\pm,\pm}^{(1)}$ and $T_{0,0}^{(2)}$ is zero on the physical space with no double occupancy and no occupied $A$ sites. Furthermore, all terms of order $U^{-2}$ should be discarded for consistency with our first expansion Eq.~\ref{appeq_EliminateU}. Using these insights, we obtain
\begin{align}
\mathcal{H}_{\rm tot}^{U \Delta} & = - \frac{1}{U} T_{0,-}^{(2)} T_{0,+}^{(2)} - \frac{1}{\Delta} T_{-,0}^{(1)} T_{+,0}^{(1)} + \frac{1}{\Delta^2} T_{-,0}^{(1)} T_{0,0}^{(2)} T_{+,0}^{(1)} \notag \\
& + \frac{1}{U\Delta} \left[ T_{-,0}^{(1)} T_{+,-}^{(1)} T_{0,+}^{(2)} + T_{0,-}^{(2)} T_{-,+}^{(1)} T_{+,0}^{(1)} \right] \notag \\
& - \frac{1}{U \Delta^2} T_{-,0}^{(1)} T_{+,-}^{(1)} T_{-,+}^{(1)} T_{+,0}^{(1)} + P
\end{align}
where we have isolated the contribution
\begin{equation}
P = \frac{1}{U\Delta^2} \left[ \frac{ T_{-,0}^{(1)} T_{+,0}^{(1)}  T_{0,-}^{(2)} T_{0,+}^{(2)} }{2} -  T_{-,0}^{(1)}  T_{0,0}^{(2)}  T_{+,-}^{(1)} T_{0,+}^{(2)} + hc \right] 
\end{equation}
that we from now on neglect since its magnitude $|t_1 t_2|^2 / (U\Delta^2)$ is much smaller than the other terms appearing in $\mathcal{H}_{\rm tot}^{U \Delta}$ when $|t_2| \ll |t_1|$, as is the case for our effective tight-binding model (see~\ref{app_wanniers}).

\begin{figure}
\centering
\includegraphics[width=\columnwidth]{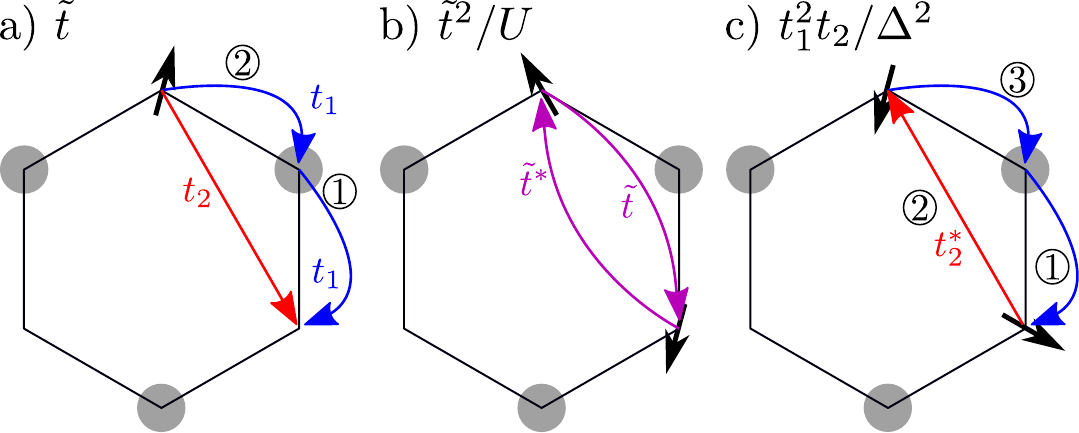}
\caption{Multi-tunneling processes appearing in the perturbation expansion for large displacement field and large interaction scale. a) Effective tunneling $\tilde{t} = t_2 + t_1^2/\Delta$ on the $B$ triangular lattice after elimination of the $A$ site with large potential energy. b) Spin exchange $\propto \tilde{t}^2/U$ between localized spins. c) Additional loop exchange $\propto t_1^2 t_2/U$ process which does not involve any double occupations. The choice of $t_2$/$t_2^*$ in the diagrams applies to spin-up electrons. Grey circles denote doubly occupied $B$ sites, while arrows represents the spin of the holes on $A$ sites.}
\label{fig:CompositeTunneling}
\end{figure}

Discarding an overall energy shift, we can rewrite the Hamiltonian derived above as
\begin{align} \label{appeq_brutspinhamiltonian}
\mathcal{H}_S = & - \frac{1}{U} \left[ T_{0,-}^{(2)} - \frac{1}{\Delta} T_{-,0}^{(1)} T_{+,-}^{(1)}  \right] \left[ T_{0,+}^{(2)} - \frac{1}{\Delta} T_{-,+}^{(1)} T_{+,0}^{(1)}  \right] \notag \\ & + \frac{1}{\Delta^2} T_{-,0}^{(1)} T_{0,0}^{(2)} T_{+,0}^{(1)} . 
\end{align}
The first line simply describes the virtual occupation of a doubly excited state using the bare or an effective next-nearest tunneling, while the second line corresponds to a particle exchange process where particle turn around a small loop of the honeycomb lattice to avoid double occupations. They are sketched in Fig.~\ref{fig:CompositeTunneling}.

It is worth noting here that the terms obtained in $\mathcal{H}_S$ are simply those one would have guessed without decoupling the $U$ and $\Delta$ expansions. This is however only due to the facts that the operators $\tilde{\mathcal{T}}_{\pm 2}$ are zero in the physical space considered in our model. This is why we have preferred the more lengthy, but more reliable, method presented here.

\subsubsection{Rewriting with spin operators}

Introducing the effective tunneling $\tilde{t} = t_2 + t_1^2/\Delta$, we can rewrite the first line of Eq.~\ref{appeq_brutspinhamiltonian} as
\begin{equation}
\frac{4}{U} \! \sum_{\langle i, j \rangle_B} \! |\tilde{t}|^2 s_i^z s_j^z + {\rm Re}\,(\tilde{t}^2) (s_i^x s_j^x + s_i^y s_j^y) + {\rm Im}\,(\tilde{t}^2) (\bm{s}_i \times \bm{s}_j) \cdot \bm{z} 
\end{equation}
which corresponds to the processes sketched in Fig.~\ref{fig:CompositeTunneling}a and~\ref{fig:CompositeTunneling}b. With a similar calculation, we can recast the second line as 
\begin{equation}
\frac{4t_1^2}{\Delta^2} \! \sum_{\langle i, j \rangle_B} \! {\rm Re}\,(t_2) \bm{s}_i \cdot \bm{s}_j + {\rm Im}\,(t_2) (\bm{s}_i \times \bm{s}_j) \cdot \bm{z} ,
\end{equation}
which is sketched in Fig.~\ref{fig:CompositeTunneling}c.
Putting everything together, we obtain an XXZ model with Dzyaloshinskii–Moriya interactions for the spin localized on the triangular $B$ lattice:
\begin{equation} \label{appeq_spinmodels}
\mathcal{H}_S = \! \sum_{\langle i, j \rangle_B} \! J_\parallel s_i^z s_j^z + J_\perp (s_i^x s_j^x + s_i^y s_j^y) + D \left[ (\bm{s}_i \times \bm{s}_j) \cdot \bm{z} \right] ,
\end{equation}
where the coefficients read
\begin{subequations} \begin{eqnarray}
J_\parallel &=& \frac{4 |\tilde{t}|^2}{U} + {\rm Re} \left( \frac{4t_1^2 t_2}{\Delta^2} \right) , \\
J_\perp &=& {\rm Re} \left( \frac{4 \tilde{t}^2}{U} +  \frac{4t_1^2 t_2}{\Delta^2} \right) , \\
D &=& {\rm Im} \left( \frac{4 \tilde{t}^2}{U} + \frac{4t_1^2 t_2}{\Delta^2} \right) . \end{eqnarray} \end{subequations}
This is the form used in the main text.

\subsubsection{Classical phase diagram}

The phases appearing for large displacement fields in our Hartree-Fock study (Fig. 5 in the main text) can be rationalized by studying the classical version of Eq.~\ref{appeq_spinmodels}. Treating the spins as classical vectors of fixed lengths and imposing the 'weak constraint' of Luttinger-Tisza~\cite{LuttingerTisza}, the spin model can be exactly solved in momentum space. 

We find spin eigenstates with spin polarized along $\bm{z}$ with the dispersion 
\begin{equation}
E_z (\bm{k}) = J_\parallel C (\bm{k}) ,
\end{equation}
and two other with spins pined in the $xy$ plane
\begin{equation}
E_{xy}^\pm (\bm{k}) = J_\perp C (\bm{k}) \pm |D S(\bm{k})| ,
\end{equation}
where we have introduced the two auxiliary functions 
\begin{equation}
C (\bm{k}) = \sum_{j=1}^3 \cos (\bm{k} \cdot \bm{a}_j) , \quad S (\bm{k}) = \sum_{j=1}^3 \sin (\bm{k} \cdot \bm{a}_j) ,
\end{equation}
with $\bm{a}_3 = -(\bm{a}_1+\bm{a}_2)$. The competition between the out of plane and in-plane modes was studied beyond the classical approximation with the ED spin-wave analysis (see main text). Thus, we discard the $E_z$ eigenstates and look for spin patterns that minimize $E_{xy}^\pm$.

Depending on $J_\perp$ and $D$, the minimum of $E_{xy}^-$ is either at the $\gamma$ point or at $\kappa_\pm$, which respectively describe the FM$_{xy}$ and AFM$_{xy}$ phases of the main text, after reintroduction of Luttinger-Tisza's 'strong constraint'~\cite{LuttingerTisza}. Comparison of the energies at these three momenta give the conditions
\begin{equation}
\begin{cases}
{\rm FM}_{xy} : & |D| + \sqrt{3} J_\perp <0 , \\
{\rm AFM}_{xy} : & |D| + \sqrt{3} J_\perp >0 .
\end{cases} 
\end{equation}
\bibliographystyle{apsrev4-2}
\bibliography{references}{}